\documentclass[twocolumn,epjc3]{svjour3}

\smartqed

\usepackage{mathptmx}
\RequirePackage[numbers,sort&compress]{natbib}
\RequirePackage[colorlinks,citecolor=blue,urlcolor=blue,linkcolor=blue]{hyperref}

\RequirePackage{graphicx}
\RequirePackage{amsmath}
\RequirePackage[cal=cm,scr=boondoxo,scrscaled=1.05]{mathalfa}  % required for lowercase script \mathscr and caligraphic \mathcal  - in the same font as is used for labels on figures (necessary)

\newcommand{\ee}{\mathrm{e}}  % upright Euler's number e

\journalname{Eur. Phys. J. C}

\begin{document}

\title{Neutrino trapping in extremely compact Tolman~VII spacetimes}

\author{Zden\v{e}k Stuchl\'{\i}k \thanksref{e1,addrA}\and 
        Jan Hlad\'{\i}k \thanksref{e2,addrA} \and 
        Jaroslav Vrba \thanksref{e3,addrA} \and 
        Camilo Posada \thanksref{e4,addrA}}

        \thankstext{e1}{e-mail: zdenek.stuchlik@physics.slu.cz \emph{(corresponding author)}}
        \thankstext{e2}{e-mail: jan.hladik@physics.slu.cz}
        \thankstext{e3}{e-mail: jaroslav.vrba@physics.slu.cz}
        \thankstext{e4}{e-mail: camilo.posada@physics.slu.cz}

\institute{Research Centre for Theoretical Physics and Astrophysics, Institute of Physics, Silesian University in Opava,\\ Bezru\v{c}ovo n\'am.~13, 746\,01 Opava, Czech Republic \label{addrA}}

% ORCID
%   Stuchlik, Z.:   0000-0003-2178-3588
%   Hladik, J.:     0000-0003-2034-4719
%   Vrba, J.:       no ORCID
%   Posada, C.:     0000-0001-8826-1974

\date{Received: date / Accepted: date}

\maketitle

\abstract{Extremely compact objects trap gravitational wa\-ves or neutrinos, assumed to move along null geodesics in the trapping regions. The trapping of neutrinos was extensively studied for spherically symmetric extremely compact objects constructed under the simplest approximation of the uniform energy density distribution, with radius located under the photosphere of the external spacetime; in addition, uniform emissivity distribution of neutrinos was assumed in these studies. Here we extend the studies of the neutrino trapping for the case of the extremely compact Tolman~VII objects representing the simplest generalization of the internal Schwarzschild solution with uniform distribution of the energy density, and the correspondingly related distribution of the neutrino emissivity that is thus again proportional to the energy density; radius of such extremely compact objects can overcome the photosphere of the external Schwarzschild spacetime. In dependence on the parameters of the Tolman~VII spacetimes, we determine the ``local'' and ``global'' coefficients of efficiency of the trapping and de\-mon\-stra\-te that the role of the trapping is significantly stron\-ger than in the internal Schwarzschild spacetimes. Our results indicate possible influence of the neutrino trapping in cooling of neutron stars.
\keywords{extremely compact object \and Tolman~VII spacetime \and trapped null geodesic \and escape cones}
\PACS{04.20.-q \and 04.40.Dg}
}

%%%%%%%%%%%%%%%%%%%%%%%%%%%%%%%%%%%%%%%%%%%%%%%%%%%%%%%%%%%%%%%%%%%%%%%%%%%%%%%%%%%%%%%%%%%%%%%%%%%%%%%%%%%%%%%%%%%%%%%%%%%%%%%%%%%%%%%%%%%%%%%%%%%%%%%%%%%%%%%%%%%%%%%%%%%%%%%%%%%%%%%%%
%%%%%%%%%%%%%%%%%%%%%%%%%%%%%%%%%%%%%%%%%%%%%%%%%%%%%%%%%%%%%%%%%%%%%%%%%%%%%%%%%%%%%%%%%%%%%%%%%%%%%%%%%%%%%%%%%%%%%%%%%%%%%%%%%%%%%%%%%%%%%%%%%%%%%%%%%%%%%%%%%%%%%%%%%%%%%%%%%%%%%%%%%

\section{Introduction}
The extremely compact objects contain a region of trapped null geodesics~\cite{Fel:69:NCBS,Abr-Mil-Stu:1993:PRD}, allowing for trapping of gravitational waves~\cite{Abr-And-Bru:1997:CQG,And:1998:APJ,KOK-SCH:1999:LRR} or neutrinos~\cite{Stu-Tor-Hle:2009:CQG}. The trapping region is centered around a stable circular null geodesic and has an outer boundary given by an unstable circular null geodesic~\cite{Stu-Tor-Hle:2009:CQG}. Extremely compact objects could be considered as black-hole mimickers in the analysis of gravitational waves detected after the merging of black holes, their mimickers, or neutron stars~\cite{Bar-etal:2019:CQG} --- the quasinormal modes of gravitational waves resulting in the merging processes are expected to be related to the unstable circular null geodesics, as demonstrated in~\cite{Car-Mir-Ber:2009:PRD}, if models based on the Einstein gravity are assumed; however, this is not necessarily true in alternative gravity theories~\cite{Kon-Stu:2017:PLB}, and exceptions are possible even in the Einstein theory combined with a non-linear electrodynamics~\cite{Stu-Sch:2019:EPJC,Tos-Stu-Sch:2018:PRD,Tos-Stu-Ahm:2019:GRQC}.

Inside the extremely compact objects representing neutron stars, or quark (hybrid) stars, null geodesics determine the motion of neutrinos, if the neutron stars are sufficiently cooled, yet maintaining large temperatures~\cite{Stu-Tor-Hle:2009:CQG,Gle:2000:BOOK,Web:1999:BOOK}. The geodesic (free) motion of the neutrinos is relevant, if their free mean path $\Lambda$ is larger than the neutron star extension $R$ which is estimated by observations to be slightly larger than $10~\mathrm{km}$. The free mean path is governed by the elastic scattering of neutrinos on electrons (neutrons), determined by the cross section $\sigma_\mathrm{e}$ ($\sigma_\mathrm{n}$), and the electron (neutron) number density $N_\mathrm{e}$ ($n_\mathrm{n}$). For neutrinos with energy $E_{\nu}$ the ne\-ut\-ri\-no--electron scattering implies the free mean path in the form~\cite{Sha-Teu:1983:BOOK}
\begin{equation}
    \Lambda_\mathrm{e} = (\sigma_\mathrm{e} N_\mathrm{e})^{-1} \sim 9 \times 10^{7} \left(\frac{\rho_\mathrm{nucl}}{\rho}\right)^{4/3} \left(\frac{100~\textrm{kV}}{E_{\nu}}\right)^3~\textrm{km}\, ,
\end{equation}
while the free mean path based on the neutrino-neutron scattering reads
\begin{equation}
    \Lambda_\mathrm{n} = (\sigma_\mathrm{n} n_\mathrm{n})^{-1} \sim  300 \left(\frac{\rho_\mathrm{nucl}}{\rho}\right) \left(\frac{100~\textrm{kV}}{E_{\nu}}\right)^2~\textrm{km}\, .
\end{equation}

We thus find $\Lambda_\mathrm{e} > 10$~km for $E_{\nu} < 20$~MeV, and $\Lambda_\mathrm{n} > 10$~km for $E_{\nu} < 500$~keV, and it is clear that in one hour old neutron star with temperature $T < 10^9$~K ($E_{\nu} < 100$~keV) the neutrino motion can be of geodesic character with good precision~\cite{Sha-Teu:1983:BOOK}.

The neutrino trapping is then relevant both for detectable decrease of the neutrino flow observed at large distances, and for significant role in the cooling of the neutron star that can influence its internal structure due to induced internal flows, causing self-organization of the neutron star matter in the trapping region. In the simplest case of the internal Schwarzschild spacetime~\cite{Sch:1916:AkadW}, the neutrino trapping was treated in detail in~\cite{Stu-Tor-Hle:2009:CQG}, and with inclusion of the cosmological constant in~\cite{Stu:2000:APS,Stu-Hla-Urb:2011:GRG}; relevance of the cosmological constant in astrophysics is discussed in~\cite{Stu-Kol-Kov:2020:UNI}. The generalization of the internal Schwarzschild spacetime in the framework of the Hartle--Thorne theory of slowly rotating compact objects~\cite{Hartle:1967he,Har:1968:APJ} was presented in~\cite{Chandra:1974MNRAS}. The trapping in the internal Schwarzschild spacetimes with inclusion of the rotational effects based on the linearized form of the Hartle--Thorne model has been recently studied in~\cite{Vrb-Urb-Stu:2020:EPJC}. Neutrino trapping in the braneworld~\cite{Ran-Sun:1999:PRL,Ger-Maa:2001:PRD} extremely compact objects with uniformly distributed energy density was treated in~\cite{Stu-Hla-Urb:2011:GRG}.

In the internal Schwarzschild spacetimes with uniform distribution of energy density~\cite{Stu-Tor-Hle:2009:CQG} the neutrino trapping is possible if its radius $R < 3GM/c^2 = 3r_\mathrm{g}/2$, where $M$ is the mass and $r_\mathrm{g}$ is the related gravitational radius of the extremely compact object, \textit{i.e.}, the object radius must be located under the unstable null circular geodesic (pho\-to\-sphe\-re) of the external vacuum Schwarzschild spacetime. The neutrino trapping efficiency increases with decreasing radius of the uniform sphere --- its radius $R\geq R_\mathrm{c}=9r_\mathrm{g}/8$~\cite{Stu-Tor-Hle:2009:CQG}. It is quite interesting that the internal Schwarzschild spacetimes demonstrate an extraordinary character for $R \to 2 M$, being related to gravastars as shown in~\cite{Kon-Pos-Stu:2019:PRD,Pos-Chi:2019:CQG,Ova-Pos-Stu:2019:CQG}.

However, from the astrophysical point of view it is important, if models of the extremely compact objects trapping neutrinos can have their radius larger than the radius of the external Schwarzschild spacetime photosphere, being thus closer to the radii of neutron stars governed by the realistic equations of state that are restricted by observations to be larger than $R = 3.2 M$ as implied by the limits of realistic equations of state applied for neutron stars. One of the well known spacetimes satisfying this requirement is represented by the special class of the solutions of Einstein gravitational equations, namely the Tolman~VII solution assuming inside the object a special, very simple but non-uniform, energy density radial profile of quadratic character~\cite{Tol:1939:PR}. This solution modifies in a realistic way the internal Schwarz\-schi\-ld solution and its trapping versions allow for radii overcoming the radius of the photosphere~\cite{Nea-Ish-Lak:2001:PRD}, making thus the extremely compact Tolman~VII models more plausible in comparison with those limited by the photosphere radius. \footnote{Note that also extremely compact polytropic spheres can have extension overcoming the external Schwarzschild spacetime photosphere radius $R = 3M$~\cite{Stu-Hle-Nov:2016:PRD,Nov-Hla-Stu:2017:PRD,Hod:2018:EPJC,Pen:2020:ARX,Hla-Pos-Stu:2020:IJMPD,Pos-Hla-Stu:2020:PRD,Peng:2020:EPJC}; for polytropic index $n > 3.3$ such extremely compact polytropes can be very extended ($R\gg r_g$) modeling thus dark matter halos of (galaxy) mass $10^{12} M_\odot$, while gravitational instability of the trapping zone of extension $R_\mathrm{tr}\ll R$ can induce gravitational collapse creating a supermassive ($M\sim10^9 M_\odot$) black hole~\cite{Stuchlik17}.} Properties of the Tolman~VII solution were studied in a series of papers~\cite{Nea-Ish-Lak:2001:PRD,Jia-Yag:2019:PRD,Jia-Yag:2020:PRD,Maurya:2019zyc,Maurya:2020rny} that all lead to a strong conclusion that this exact solution of the Einstein equations exhibits surprisingly good approximation to properties of realistic neutron stars. Furthermore, a modified Tolman~VII solution was introduced in~\cite{Jia-Yag:2019:PRD} that includes an additional quartic term in the energy density radial profile being, however, only an approximate solution of the Einstein equations; its concordance with realistic models of neutron stars was discussed and confirmed along the I--Love--C relations~\cite{Jia-Yag:2020:PRD}. Moreover, also an anisotropic version of the Tolman~VII solution was presented in~\cite{Hen-Stu:2019:EPJC} enabling inclusion of the influence of additional matter sources on the properties of neutron stars. Here we restrict attention to the study of the neutrino trapping effect in Tolman~VII solutions, as it is an exact solution of the Einstein equations giving very good approximation to the behavior of the realistic neutron stars~\cite{Hen-Stu:2019:EPJC}.

In Sect.~\ref{SECTION2} we summarize the general properties of the Tolman~VII spacetimes, and by treating the null geodesics of these spacetimes we determine the range of parameters (external radii $R$) of the extremely compact Tolman~VII spacetimes allowing for existence of the region of neutrino trapping. Then we discuss in Sect.~\ref{SECTION3} the effective potential and trapping (or complementary escape) cones of the null geo\-de\-sics. In Sect.~\ref{SECTION4} we determine the ``local'' and ``global'' efficiency coefficients of the null geodesics (neutrino) trapping, and make comparison to the simplest case of the internal Schwarz\-schild spacetimes. Concluding remarks are presented in Sect.~\ref{SECTION5}.

In the following we use geometric units with $c = G = 1$.

%%%%%%%%%%%%%%%%%%%%%%%%%%%%%%%%%%%%%%%%%%%%%%%%%%%%%%%%%%%%%%%%%%%%%%%%%%%%%%%%%%%%%%%%%%%%%%%%%%%%%%%%%%%%%%%%%%%%%%%%%%%%%%%%%%%%%%%%%%%%%%%%%%%%%%%%%%%%%%%%%%%%%%%%%%%%%%%%%%%%%%%%%
\section{Tolman~VII spacetime and its null geodesics}\label{SECTION2}
In the static and spherically symmetric Tolman~VII solution, the energy density distribution is assumed to be a quadratic function of the radius, and the Einstein equations and the stress--energy conservation then enable the determination of the metric coefficients and pressure radial profiles in terms of elementary functions.

We first present the Tolman~VII solution as given in an elegant and compact form in~\cite{Jia-Yag:2019:PRD} where the free parameter characterizing the solution is chosen to be the compactness of the object $\mathcal{C} \equiv M/R$, or its inverse, $R/M$. Then we discuss the equations of its null geodesics and give the extremely compact Tolman~VII spacetimes allowing for the trapping of null geodesics by determining the limits on the values of the parameter $R/M$.

\subsection{Tolman~VII spacetime}
The Tolman~VII spacetime belongs to the static and spherically symmetric spacetimes having in the standard Schwarz\-schild coordinates ($t,r,\theta,\varphi$) the line element in the form
\begin{equation}\label{met}
    \mathrm{d}s^2 = -\ee^{\Phi(r)}\mathrm{d}t^2+\ee^{\Psi(r)}\mathrm{d}r^2 + r^2\left(\mathrm{d}\theta^2 + \sin^2\theta \, \mathrm{d}\varphi^2\right)\, .
\end{equation}
The matter inside the Tolman~VII solution is assumed to be a perfect fluid with stress--energy tensor
\begin{equation}\label{set}
    T_{\mu\nu} = (\rho + p)u_{\mu}u_{\nu} + pg_{\mu\nu}\, ,
\end{equation}
where $\rho$ is the energy density and $p$ is the pressure of the fluid. The metric functions, the energy density and pressure are functions of the radius $r$ only. The Einstein equations then give a set of three differential equations for the four unknown functions of the radius~$r$~\cite{Tol:1939:PR}:
\begin{multline}
    \frac{\mathrm{d}}{\mathrm{d}r}\left(\frac{\ee^{-\Psi}-1}{r^2}\right) + \frac{\mathrm{d}}{\mathrm{d}r}\left(\frac{\ee^{-\Psi}\Phi'}{2r}\right) \\
        + \ee^{-\Psi-\Phi}\frac{\mathrm{d}}{\mathrm{d}r}\left(\frac{\ee^{-\Phi}\Phi'}{2r}\right) = 0\, ,
\end{multline}
\begin{equation}
    \ee^{-\Psi}\left(\frac{\Phi'}{r}+\frac{1}{r^2}\right) - \frac{1}{r^2} = 8\pi p\, ,
\end{equation}
\begin{equation}
    \ee^{-\Psi}\left(\frac{\Psi'}{r}-\frac{1}{r^2}\right) + \frac{1}{r^2}=8\pi \rho\, ,
\end{equation}
where the prime denotes derivative against $r$, and the radial metric coefficient is given directly in terms of the mass contained inside the radius $m(r)$  in the form
\begin{equation}\label{mr}
    \ee^{-\Psi(r)} = 1 - \frac{2m(r)}{r}\, .
\end{equation}
Usually, this system of equations is closed after the specification of the equation of state $p(\rho)$, however, the Tolman~VII solution is governed by specification (assumption) of the energy density radial profile~\cite{Tol:1939:PR}. The exterior at $r > R$, where both energy density and pressure vanish, is described by the standard Schwarzschild metric
\begin{equation}
    \ee^{\Phi_\mathrm{ex}} = \ee^{-\Psi_\mathrm{ex}} = 1 - \frac{2M}{r}\, ,
\end{equation}
with the total mass of the object $M = m(R)$.

In the Tolman~VII solution the energy density profile, given in terms of the dimensionless radial coordinate $\xi = r/R$, is given by the relation
\begin{equation}
    \rho(\xi) = \rho_\mathrm{c}(1 - \xi^2)\, ,
\end{equation}
where $R$ denotes radius of the object, and $\rho_\mathrm{c}$ is the central energy density. Notice that the energy density vanishes at the surface \footnote{Here we are studying extremely compact objects, for which the zero density at the surface is a good approximation considering the interior density profile, as the central energy density is several orders of magnitude higher than the surface energy density.}. 
We can then immediately obtain the mass radial profile which takes the form
\begin{equation}
    m(r) = 4\pi\rho_\mathrm{c}\left(\frac{r^3}{3} - \frac{r^5}{5R^2}\right) \, .
\end{equation}
Using the compactness parameter $\mathcal{C} \equiv M/R$, the central density can be expressed in the form
\begin{equation}
    \rho_\mathrm{c} = \frac{15\mathcal{C}}{8\pi R^2} \, .
\end{equation}
Then the Tolman~VII solution can be expressed in the form introduced by~\cite{Jia-Yag:2019:PRD,Jia-Yag:2020:PRD}. The energy density and mass profiles read
\begin{equation}\label{eq13}
    \rho(\xi) = \frac{15\mathcal{C}}{8\pi R^2}(1 - \xi^2)\, ,
\end{equation}
\begin{equation}
    m(\xi) = \mathcal{C} R\left(\frac{5}{2}\xi^3 - \frac{3}{2}\xi^5\right)\, .
\end{equation}
The metric coefficients are determined by the relations
\begin{equation}
    \ee^{-\Psi(\xi)} = 1 - \mathcal{C}\xi^2(5 - 3\xi^2)\, ,
\end{equation}
\begin{equation}
    \ee^{\Phi(\xi)} = C_1 \cos^2\phi_\mathrm{T}\, ,
\end{equation}
while the pressure is given by
\begin{equation}
    p(\xi) = \frac{1}{4\pi R^2}\left[\sqrt{3\mathcal{C}\ee^{-\Psi}}\tan\phi_\mathrm{T} - \frac{\mathcal{C}}{2}(5 - 3\xi^2)\right]\, ,
\end{equation}
where
\begin{equation}
    \phi_\mathrm{T}(\xi) = C_2 - \frac{1}{2}\log\left(\xi^2 - \frac{5}{6} + \sqrt\frac{\ee^{-\Psi}}{3\mathcal{C}}\right)\, ,
\end{equation}
and the integration constants take the form
\begin{equation}
    C_1 = 1 - \frac{5\mathcal{C}}{3}\, ,
\end{equation}
\begin{equation}
    C_2 = \arctan\sqrt\frac{\mathcal{C}}{3(1-2\mathcal{C})} + \frac{1}{2}\log\left(\frac{1}{6} + \sqrt\frac{1 - 2\mathcal{C}}{3\mathcal{C}}\right)\, .
\end{equation}
Properties of the Tolman~VII solution were extensively studied in several papers --- the existence of the trapping null geodesics~\cite{Nea-Ish-Lak:2001:PRD}, stability against radial per\-tur\-ba\-ti\-ons~\cite{Negi:1999GR,Negi:2001as,Rag-Hob:2015:PRD,Moustakidis:2016ndw}, and other important properties relevant for comparison to the realistic neutron stars were discussed recently in~\cite{Jia-Yag:2019:PRD,Jia-Yag:2020:PRD}. Here we shortly summarize the most relevant results.

The well defined Tolman~VII spacetimes are restricted by the requirement of finite central pressure $p(\xi = 0)$. Clearly, the central pressure diverges if $\tan\phi_\mathrm{T}\,(\xi = 0)$ diverges, \textit{i.e.}, if
\begin{equation}
    \phi_\mathrm{T}\,(\xi=0)=\frac{\pi}{2}\, .
\end{equation}
This condition then implies the relation
\begin{equation}
    \frac{\pi}{2} = C_2 - \frac{1}{2} \log\left(\sqrt{\frac{1}{3\mathcal{C}}} - \frac{5}{6}\right)\, .
\end{equation}
We then find the lower limiting value of the Tolman~VII sphere to be given as
\begin{equation}
    R_\mathrm{T\, min} \doteq 2.5894 M\, .
\end{equation}
We can see that this critical minimal radius of the Tolman VII spheres is substantially exceeding the lowest radius allowed for the internal Schwarzschild solutions with u\-ni\-form energy density that reads $R_\mathrm{S\, min} =  2.25M$ \cite{Stu-Tor-Hle:2009:CQG}.

In order to find the class of extremely compact Tol\-man VII solution that allows the trapping of null geodesics, we have to study the null geodesics of these spacetimes. %Clearly, $R_\mathrm{T\, min}$ is the lowest radius of the extremely compact Tolman~VII solutions.

\subsection{Circular null geodesics in extremely compact Tolman~VII spacetimes}
The motion along null geodesics is governed by the geodesic equation and corresponding normalization condition
\begin{equation}\label{geod}
    \frac{\mathrm{D}p^{\mu}}{\mathrm{d}\tau}=0\, ,\qquad p_{\mu}p^{\mu}=0\, ,
\end{equation}
where $\tau$ is the affine parameter. Two Killing vector fields ($\partial/\partial t\,, \partial/\partial \varphi$)  imply two conserved components of the four-momentum
\begin{equation}
\begin{aligned}\label{const}
    p_t &= -E & \qquad &\textrm{(energy)}\, ,\\
    p_{\varphi} &= \phi & &\textrm{(axial angular momentum)}\, .
\end{aligned}
\end{equation}
Motion in the spherically symmetric spacetimes is restricted to their central planes. In the case of geodesic motion, it is convenient to choose the equatorial plane of the coordinate system, \textit{i.e.} we set $\theta=\pi/2$.
Introducing the impact pa\-ra\-me\-ter $\lambda=\phi/E$, we obtain from the normalization condition relation governing the radial motion in the form
\begin{equation}
    (p^r)^2 = \ee^{-\left(\Phi+\Psi\right)}\, E^2\left(1 - \ee^{\Phi} \frac{\lambda^2}{r^2}\right)\, .
\end{equation}
It is obvious that for null geodesics the energy $E$ is not relevant for the character of the motion (we can use it for the scaling of the impact parameter $\lambda$). Note that the expression in brackets is non-negative. We can thus introduce the effective potential $\mathrm{V}_\mathrm{eff}$ determining the turning points of the radial motion along null geodesics for a given impact parameter~$\lambda$~\cite{Mis-Tho-Whe:1973:BOOK}:
\begin{multline}\label{e:sveff}
    \lambda^2\leq V_\mathrm{eff} = \\
        \begin{cases}
            V{}_\mathrm{eff}^\mathrm{int} = \displaystyle\frac{3r^2}{3-5\mathcal{C}}\cos ^{-2}\left[C_\mathrm{a} + Y(r)\right]\, , &\text{for } r \le R\, ,\\[10pt]
            V{}_\mathrm{eff}^\mathrm{ext} = \displaystyle\frac{r^3}{r - 2M}\, , &\text{for }  r > R\, ,
        \end{cases}
\end{multline}
where
\begin{equation}
    C_\mathrm{a} = \arctan\sqrt{\frac{\mathcal{C}}{3 - 6\mathcal{C}}}\, ,
\end{equation}
\begin{multline}
    Y(r) = \\ \frac{1}{2}\log\left[\frac{\left(1 + 2 \sqrt{3/\mathcal{C} - 6}\right) R^2}{2 \sqrt{3R^4/\mathcal{C} + 9 r^4 - 15 r^2R^2}+6 r^2 - 5R^2}\right]\, .
\end{multline}

%%%%%%%%%%%%%%%%%%%%%%%%%%%%%%%%%%%%%%%%%%%%%%%%%%%%%%%%%%%%%%%%%%%%%%%%%%%%%%%%%%%%%%%%%%%%%%%%%%%%%%%%%%%%%%%%%%%%%%%%%%%%%%%%%%%%%%%%%%%%%%%%%%%%%%%%%%%%%%%%%%%%%%%%%%%%%%%%%%%%%%%%%
\begin{figure}[t]
	\centering\includegraphics[width=\linewidth]{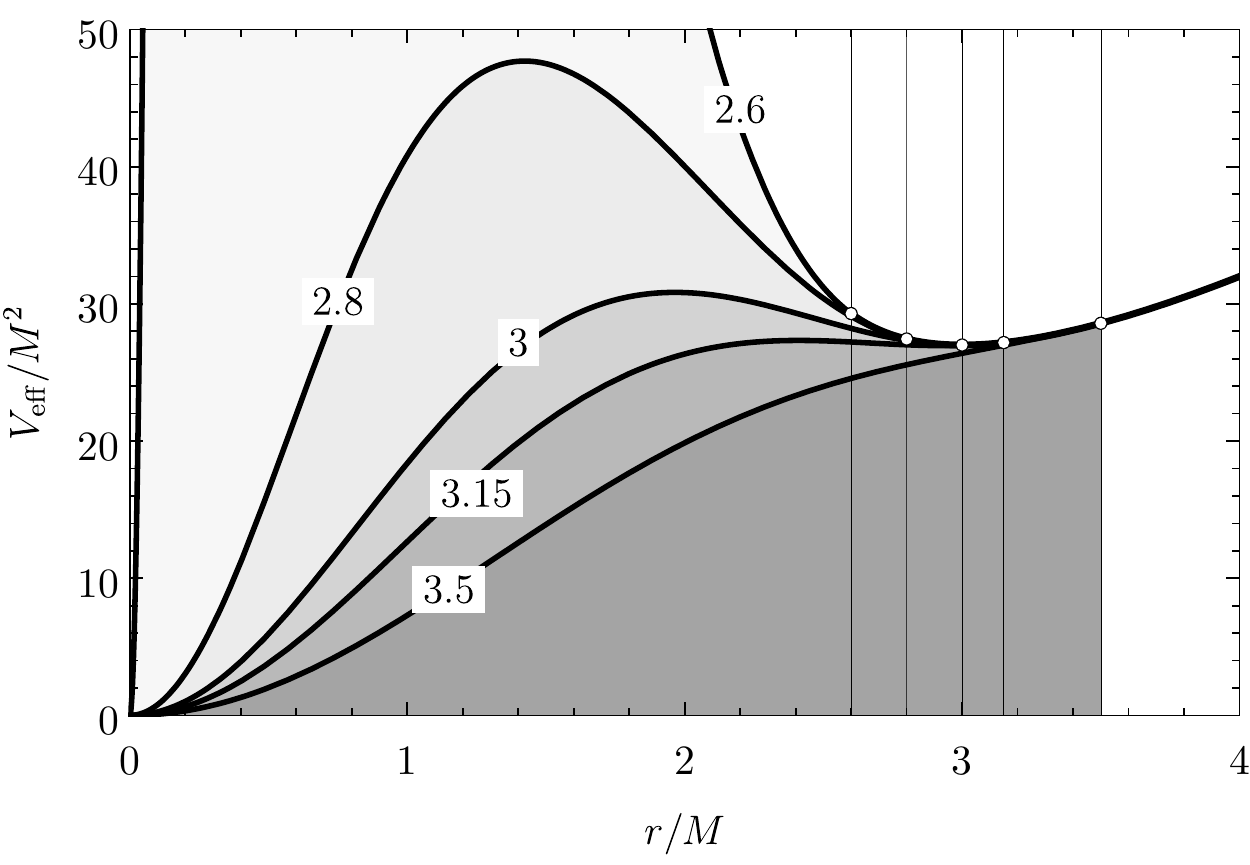}
	\caption{\label{figure1} Effective potentials of the Tolman VII objects with inverse compactness $R/M$ given in the potentials.}
\end{figure}
%%%%%%%%%%%%%%%%%%%%%%%%%%%%%%%%%%%%%%%%%%%%%%%%%%%%%%%%%%%%%%%%%%%%%%%%%%%%%%%%%%%%%%%%%%%%%%%%%%%%%%%%%%%%%%%%%%%%%%%%%%%%%%%%%%%%%%%%%%%%%%%%%%%%%%%%%%%%%%%%%%%%%%%%%%%%%%%%%%%%%%%%%

The behavior of the resulting effective potentials is presented in Fig.~\ref{figure1}, where one can notice that contrary to the case of the internal Schwarzschild spacetimes, the effective potential of the interior of the Tolman~VII spacetimes has a non-monotonic character even for $R > 3M$.

%%%%%%%%%%%%%%%%%%%%%%%%%%%%%%%%%%%%%%%%%%%%%%%%%%%%%%%%%%%%%%%%%%%%%%%%%%%%%%%%%%%%%%%%%%%%%%%%%%%%%%%%%%%%%%%%%%%%%%%%%%%%%%%%%%%%%%%%%%%%%%%%%%%%%%%%%%%%%%%%%%%%%%%%%%%%%%%%%%%%%%%%%
\begin{figure*}[t]
	\centering
	\includegraphics[width=0.48\linewidth]{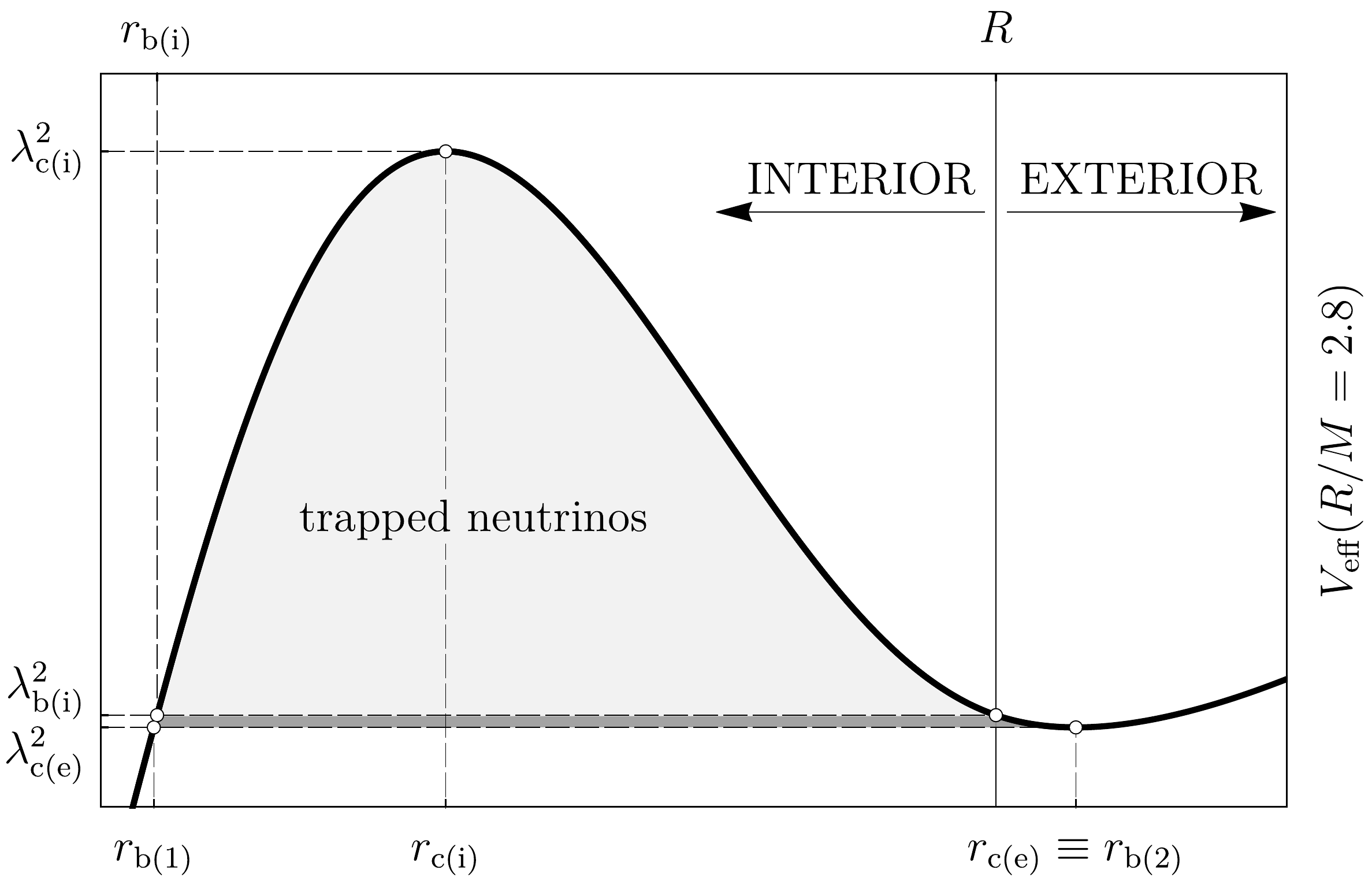}\hfill\includegraphics[width=0.48\linewidth]{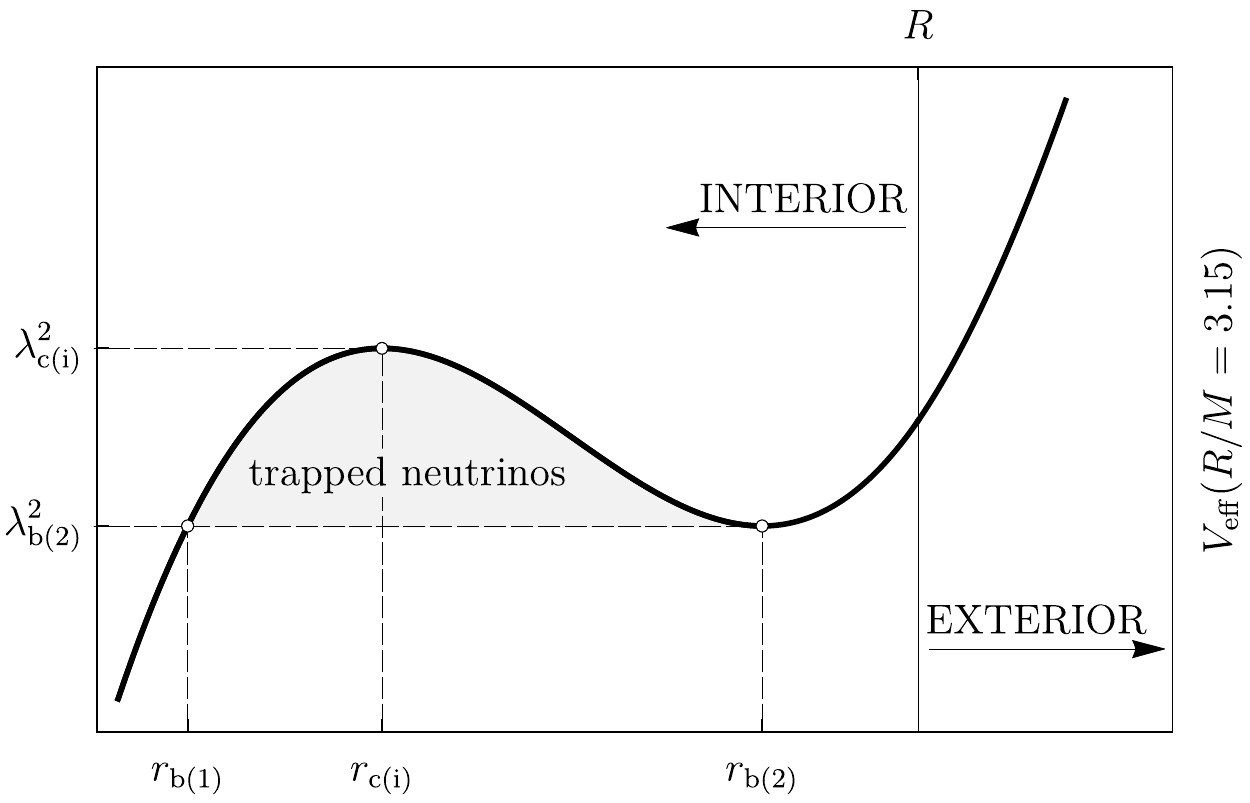}
	\caption{\label{figure2}The effective potential of the Tolman~VII spacetimes with $R=2.8 M$ (left panel) and $R=3.15 M$ (right panel). The trapped area is light shaded for the internal trapped null geodesics and dark shaded for the external trapped geodesics.}
\end{figure*}
%%%%%%%%%%%%%%%%%%%%%%%%%%%%%%%%%%%%%%%%%%%%%%%%%%%%%%%%%%%%%%%%%%%%%%%%%%%%%%%%%%%%%%%%%%%%%%%%%%%%%%%%%%%%%%%%%%%%%%%%%%%%%%%%%%%%%%%%%%%%%%%%%%%%%%%%%%%%%%%%%%%%%%%%%%%%%%%%%%%%%%%%%

The circular null geodesics are determined by the local extrema of the effective potential, \textit{i.e.}, for the interior by the condition $\mathrm{d}V{}_\mathrm{eff}^\mathrm{int}/\mathrm{d}r = 0$ that implies the relation
\begin{equation}
    \frac{6 r \left(1 + r Y'(r) \tan \chi \right)}{\left(5 \mathcal{C} - 3\right)\cos^2 \chi} = 0\, ,
    \label{e:dvef}
\end{equation}
where
\begin{equation}
   \chi = C_\mathrm{a} + Y(r)\, .
\end{equation}
The limiting case when the effective potential has an inflexion point and starts to be monotonic is determined by the additional condition $\mathrm{d}^2 V{}_\mathrm{eff}^\mathrm{int}/\mathrm{d}r^2=0$ which implies the relation
\begin{multline}
    r\left[ 2 r Y'(r)^2 \left(\cos (2 \chi)-2\right) - r Y''(r) \sin (2 \chi)\right.\\
         \left. - 4 Y'(r) \sin (2 \chi)\right]\Big/\cos^{2}{\chi} = 2\, .
\end{multline}
By numerical methods we are able to find the critical maximal value of $R$ allowing for existence of a trapping region in the Tolman~VII solution which reads
\begin{equation}
    R_\mathrm{t} \doteq 3.202 M \, .
\end{equation}
The range of radii $R_\mathrm{T\, min}\leq R \leq R_\mathrm{t}$ determines the trapping Tolman~VII spacetimes. For all of them, a local maximum of $V{}_\mathrm{eff}^\mathrm{int}$ corresponding to a stable circular null geodesic always exists, being located at $r_\mathrm{c(i)}$, with value giving the upper limit on the impact parameters of trapped null geodesics $\lambda^2_\mathrm{c(i)}=\lambda(r_\mathrm{c(i)})^2 = V{}_\mathrm{eff}^\mathrm{int}(r_\mathrm{c(i)})$.

A local minimum of the external effective potential $V{}_\mathrm{eff}^\mathrm{ext}$ corresponding to the unstable circular null geodesics in the vacuum Schwarzschild spacetime, may exist if and only if $R < 3 M$ and comes from solution $\mathrm{d}V{}_\mathrm{eff}^\mathrm{ext}/\mathrm{d}r = 0$ which gives the well known results $r_\mathrm{c(e)} = 3 M$ and $\lambda_\mathrm{c(e)}^2 = 27 M^2$~\cite{Mis-Tho-Whe:1973:BOOK}. The behavior of the effective potential of the Tolman~VII spacetime has then similar character as in the case of the internal Schwarzschild spacetimes, as demonstrated in Fig.~\ref{figure2}. In this case the (shaded) region of trapped null geodesics can be separated into two parts --- see \textit{e.g.}~\cite{Stu-Tor-Hle:2009:CQG}. The internal trapped null geodesics (light shade) have motion fully restricted to the interior of the object, being limited by the values of the impact parameters, $\lambda_\mathrm{b(i)}< \lambda < \lambda_\mathrm{c(i)}$ and the radii $r_\mathrm{b(i)} < r < R$; the external trapped null geodesics (dark shade) leave and re-enter the object, being limited by the values of the impact parameters $\lambda_\mathrm{c(e)} < \lambda < \lambda_\mathrm{b(i)}$ and the radii $r_\mathrm{b(1)} < r < r_\mathrm{c(e)}=3 M \equiv r_\mathrm{b(2)}$ --- see Fig.~\ref{figure2} (left panel). The critical impact parameters (radii) $\lambda_\mathrm{b(1)}$ ($r_\mathrm{b(1)}$) and $\lambda_\mathrm{b(i)}$ ($r_\mathrm{b(i)}$) are determined by the equations $\lambda^2_\mathrm{b(1)} \equiv V{}_\mathrm{eff}^\mathrm{int}(r = r_\mathrm{b(1)}) = V{}_\mathrm{eff}^\mathrm{ext}(r_\mathrm{c(e)}) \equiv 27 M^2$ and $\lambda^2_\mathrm{b(i)} \equiv V{}_\mathrm{eff}^\mathrm{int}(r = r_\mathrm{b(i)}) = V{}_\mathrm{eff}^\mathrm{ext}(R)$. However, in the following we are not discussing separately the internal and external trapped null geodesics, considering them only in unity.

In the case $3 M < R < R_\mathrm{t}$, the Tolman~VII spacetime has a slightly different character in comparison with the internal Schwarzschild spacetime, as its effective potential $V{}_\mathrm{eff}^\mathrm{int}(r)$ demonstrates along with presence of the local maximum giving the stable circular null geodesics at $r_\mathrm{c(i)}$ also a local minimum giving an unstable circular null geodesic that is located at the radius $r_\mathrm{b(2)}$ and has the impact parameter given by the relation $\lambda^2_\mathrm{b(2)} \equiv \lambda(r = r_\mathrm{b(2)})^2 = V{}_\mathrm{eff}^\mathrm{int}(r_\mathrm{b(2)})$ taken at the local minimum of the effective potential. Therefore, the trapping Tolman~VII spacetimes with $R > 3M$ have only internal trapped null geodesics that are limited by the impact parameters $\lambda_\mathrm{b(2)} < \lambda < \lambda_\mathrm{c(i)}$, while the range of radii where the null geodesic trapping occurs is limited by $r_\mathrm{b(1)} < r < r_\mathrm{b(2)}$. The critical radius $r_\mathrm{b(1)}$, giving the lower limit on radius of trapped null geodesics, is determined by the relation $\lambda^2_\mathrm{b(2)} = V{}_\mathrm{eff}^\mathrm{int}(r_\mathrm{b(1)})$.

\begin{figure}[t]
	\centering\includegraphics[width=\linewidth]{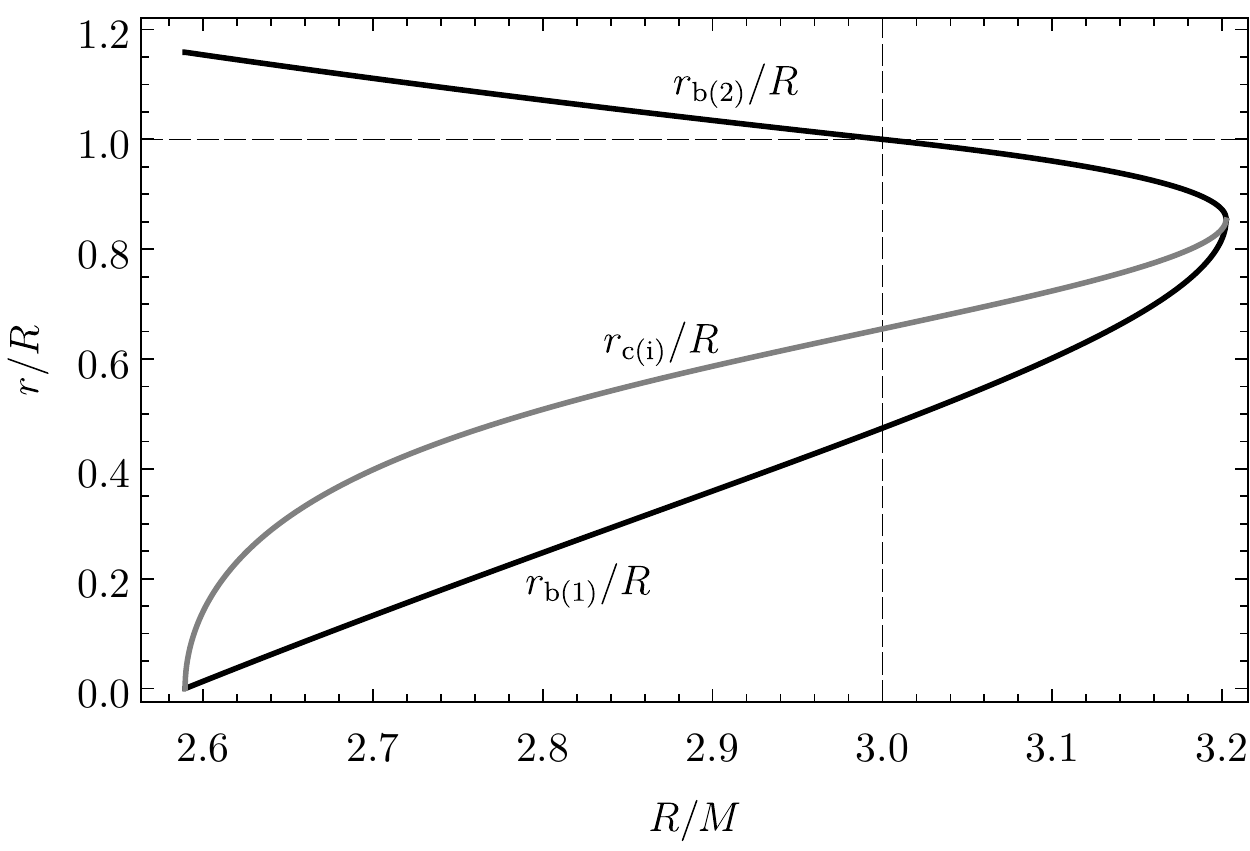}
	\caption{\label{figure3}Region of trapping of null geodesics in extremely compact Tolman~VII spacetimes. Significant radii $r_\mathrm{b(1)}$, $r_\mathrm{b(2)}$ and $r_\mathrm{c(i)}$ governing the trapping region are given in relation to the parameter $R/M$ of the object. The profiles are related to the object radius $R$. Note that for objects with $R/M < 3$ there is $r_\mathrm{b(2)}/R > 1$ as the unstable circular null geodesic is located outside the object, while it is inside for $R/M > 3$.}
\end{figure}

Contrary to the case of the internal Schwarzschild spacetime where the critical values of the impact parameters and radii governing the trapping effects can be given in terms of elementary functions, for the trapping Tolman~VII spacetimes they can be determined numerically only. The results of the numerical calculations for the critical radii are presented in Fig.~\ref{figure3}. Note that in the limiting case of $R = R_\mathrm{t}$, the photon circular null geodesics coincide at $r\doteq 0.855\,R$. The range of integration of the trapping is determined by the radii $r_\mathrm{b(1)}$ and $r_\mathrm{b(2)}$ for the spacetimes with $R > 3 M$, and by $r_\mathrm{b(1)}$ and $r_\mathrm{b(2)} = R$ in the spacetimes with $R \leq 3 M$.

%%%%%%%%%%%%%%%%%%%%%%%%%%%%%%%%%%%%%%%%%%%%%%%%%%%%%%%%%%%%%%%%%%%%%%%%%%%%%%%%%%%%%%%%%%%%%%%%%%%%%%%%%%%%%%%%%%%%%%%%%%%%%%%%%%%%%%%%%%%%%%%%%%%%%%%%%%%%%%%%%%%%%%%%%%%%%%%%%%%%%%%%%
%%%%%%%%%%%%%%%%%%%%%%%%%%%%%%%%%%%%%%%%%%%%%%%%%%%%%%%%%%%%%%%%%%%%%%%%%%%%%%%%%%%%%%%%%%%%%%%%%%%%%%%%%%%%%%%%%%%%%%%%%%%%%%%%%%%%%%%%%%%%%%%%%%%%%%%%%%%%%%%%%%%%%%%%%%%%%%%%%%%%%%%%%

\begin{figure}[t]
	\centering\includegraphics[width=\linewidth]{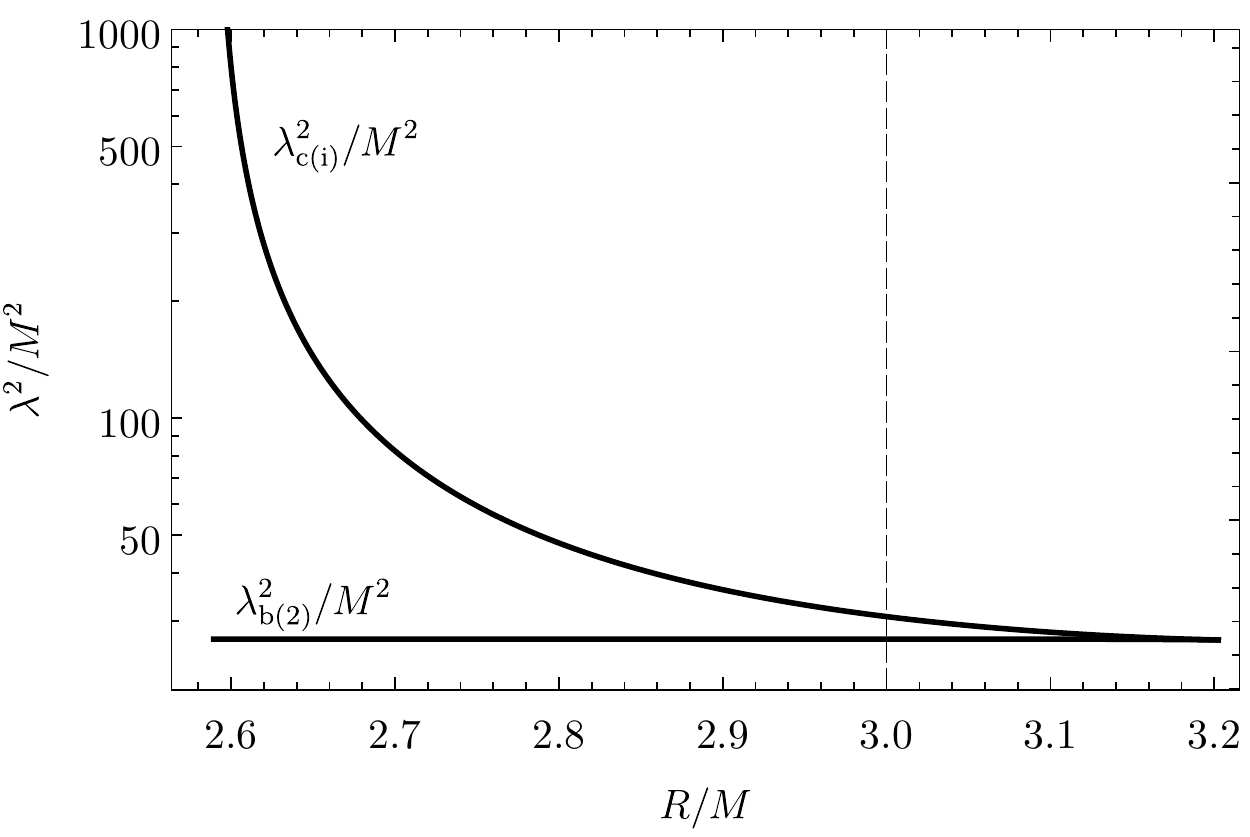}
	\caption{\label{figure4}Impact parameters governing the trapping cones of null geodesics in Tolman~VII spacetimes. Dependence of the critical values of the impact parameter, $\lambda_\mathrm{c(i)}$ and $\lambda_\mathrm{b(2))}$ is given as a function of the parameter $R/M$.}
\end{figure}

For the critical values of the impact parameter  $\lambda$ of the null geodesics, relevant for the null geodesics trapping, the dependence on the parameter $R/M$ is presented in Fig.~\ref{figure4}. The range of the values of $\lambda_\mathrm{b(2)}$ and $\lambda_\mathrm{c(i)}$ governs the efficiency of the trapping effects.

%%%%%%%%%%%%%%%%%%%%%%%%%%%%%%%%%%%%%%%%%%%%%%%%%%%%%%%%%%%%%%%%%%%%%%%%%%%%%%%%%%%%%%%%%%%%%%%%%%%%%%%%%%%%%%%%%%%%%%%%%%%%%%%%%%%%%%%%%%%%%%%%%%%%%%%%%%%%%%%%%%%%%%%%%%%%%%%%%%%%%%%%%
\section{Escape cones and trapped null geodesics}\label{SECTION3}
We now have to apply the framework of trapped null geo\-de\-sics in the context of the models of neutrinos radiated by matter of the extremely compact object. Thus, we have to determine the escape (or complementary trapping) cones of null geodesics related to the matter constituting the configuration --- the trapped part of neutrinos radiated by a given source corresponds to the directional angles belonging to the trapping cone of null geodesics related to the source.

As the Tolman~VII spacetimes are spherically symmetric and static, we can directly follow the procedures introduced in~\cite{Stu-Tor-Hle:2009:CQG} for the interior Schwarzschild solution. The escape (trapping) cones have to be related to the static sources (observers) in the static spacetime --- we have to determine the trapping cones in appropriately chosen local frames of the static observers. In a spherically symmetric spacetime the tetrad of the differential forms reads
\begin{equation}
  \begin{aligned}
	e^{(t)} &= \mathrm e^{\Phi/2}\, \mathrm{d} t\, , \quad & e^{(r)} &= \mathrm e^{\Psi/2}\, \mathrm{d} r\, ,\\
    e^{(\theta)} &= r\, \mathrm{d} \theta\, , & e^{(\varphi)} &= r\, \mathrm{sin}\theta \, \mathrm{d} \varphi\, ,
  \end{aligned}
\end{equation}
and its line element then can be expressed in the special-relativistic form
\begin{equation}
	\mathrm{d} s^2 = - \left[ e^{(t)}\right]^2 + \left[ e^{(r)}\right]^2 + \left[ e^{(\theta)}\right]^2 + \left[ e^{(\varphi)}\right]^2\, .
\end{equation}
The complementary tetrad of base 4-vectors $e_{(\alpha)}$ is determined by
\begin{equation}
	e_{(\alpha)}^{\mu}e_\nu^{(\alpha)}=\delta^\mu_\nu\, ,\qquad e_\mu^{(\alpha)}\ee^\mu_{(\beta)}=\delta^\alpha_\beta\, .
\end{equation}
Physically relevant projections of a neutrino (null geodesic) 4-momentum $p^{\mu}$ are given by
\begin{equation}
	p^{(\alpha)}=p^{\mu}e_\mu^{(\alpha)}\, ,\qquad p_{(\alpha)} = p_\mu \ee^\mu_{(\alpha)}\, .
\end{equation}
Neutrinos radiated locally by a static source can be characterized by the directional angles ($\alpha$, $\beta$, $\gamma$) related to the location of the source, whose definition is presented for instance in~\cite{Stu-Sch:2010:CQG,Stu-Cha-Sce:2018:EPJC}. These directional angles are connected by the relation
\begin{equation}
    \cos\gamma=\sin\beta\, \sin\alpha\, .
\end{equation}
Due to the spherical symmetry of the configuration, the directional angle $\alpha$ related to the radial direction (the outgoing radial unit vector $\mathbf{e}_{(r)}$) is sufficient to determine the escape (trapping) cones, while the angle $\beta$ determines position on the trapping cone; the angle $\gamma$ is related to the axial unit vector~\cite{Stu-Tor-Hle:2009:CQG}.

%%%%%%%%%%%%%%%%%%%%%%%%%%%%%%%%%%%%%%%%%%%%%%%%%%%%%%%%%%%%%%%%%%%%%%%%%%%%%%%%%%%%%%%%%%%%%%%%%%%%%%%%%%%%%%%%%%%%%%%%%%%%%%%%%%%%%%%%%%%%%%%%%%%%%%%%%%%%%%%%%%%%%%%%%%%%%%%%%%%%%%%%%
%%%%%%%%%%%%%%%%%%%%%%%%%%%%%%%%%%%%%%%%%%%%%%%%%%%%%%%%%%%%%%%%%%%%%%%%%%%%%%%%%%%%%%%%%%%%%%%%%%%%%%%%%%%%%%%%%%%%%%%%%%%%%%%%%%%%%%%%%%%%%%%%%%%%%%%%%%%%%%%%%%%%%%%%%%%%%%%%%%%%%%%%%

%%%%%%%%%%%%%%%%%%%%%%%%%%%%%%%%%%%%%%%%%%%%%%%%%%%%%%%%%%%%%%%%%%%%%%%%%%%%%%%%%%%%%%%%%%%%%%%%%%%%%%%%%%%%%%%%%%%%%%%%%%%%%%%%%%%%%%%%%%%%%%%%%%%%%%%%%%%%%%%%%%%%%%%%%%%%%%%%%%%%%%%%%
\begin{figure}[b]
	\centering
	\centering\includegraphics[width=\linewidth]{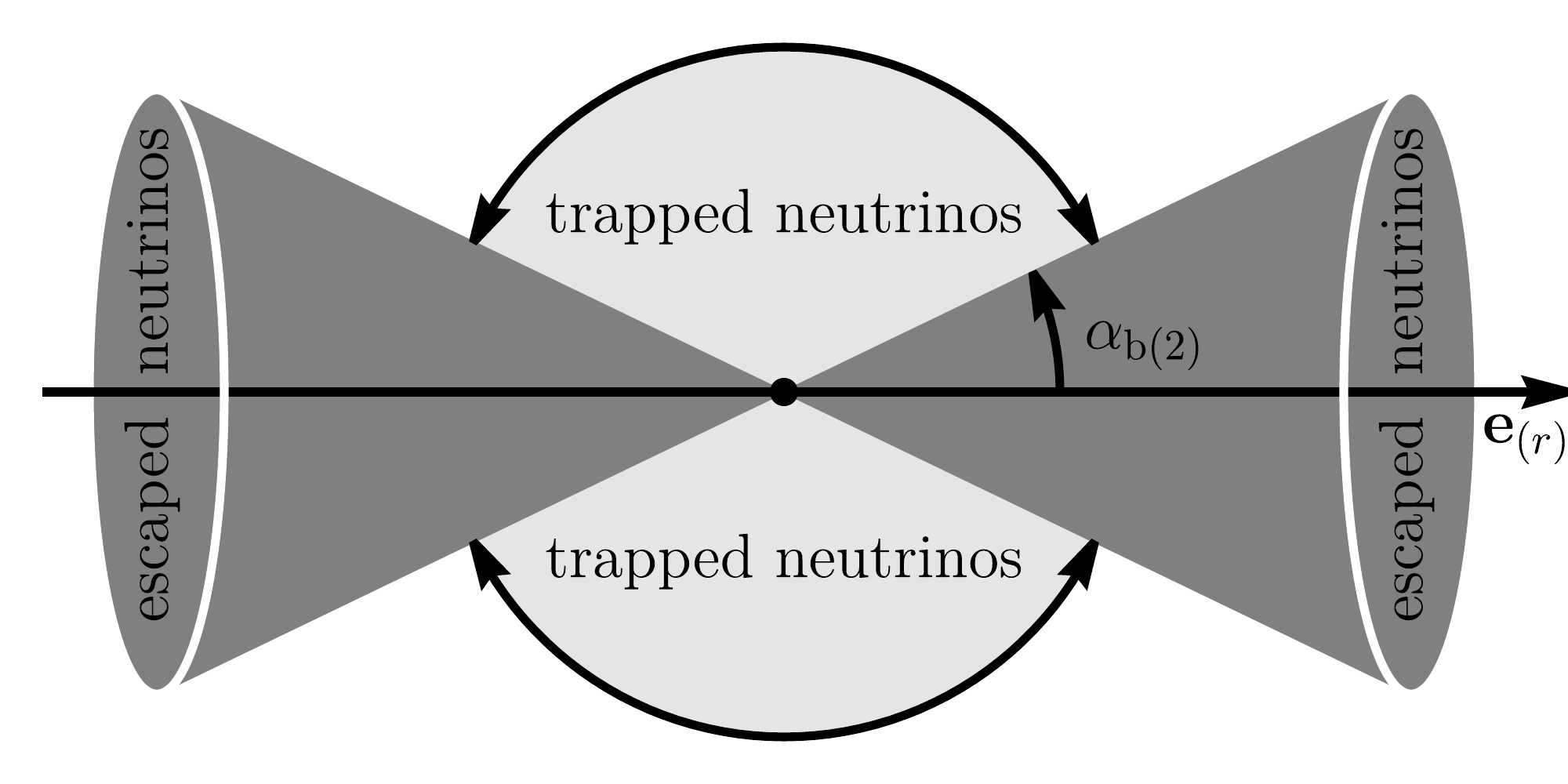}
	\caption{\label{figure5}Visualization of the escaping/trapping cone.}
\end{figure}
%%%%%%%%%%%%%%%%%%%%%%%%%%%%%%%%%%%%%%%%%%%%%%%%%%%%%%%%%%%%%%%%%%%%%%%%%%%%%%%%%%%%%%%%%%%%%%%%%%%%%%%%%%%%%%%%%%%%%%%%%%%%%%%%%%%%%%%%%%%%%%%%%%%%%%%%%%%%%%%%%%%%%%%%%%%%%%%%%%%%%%%%%

%%%%%%%%%%%%%%%%%%%%%%%%%%%%%%%%%%%%%%%%%%%%%%%%%%%%%%%%%%%%%%%%%%%%%%%%%%%%%%%%%%%%%%%%%%%%%%%%%%%%%%%%%%%%%%%%%%%%%%%%%%%%%%%%%%%%%%%%%%%%%%%%%%%%%%%%%%%%%%%%%%%%%%%%%%%%%%%%%%%%%%%
\begin{figure*}[t]
	\begin{center}
	\includegraphics[width=0.50\linewidth]{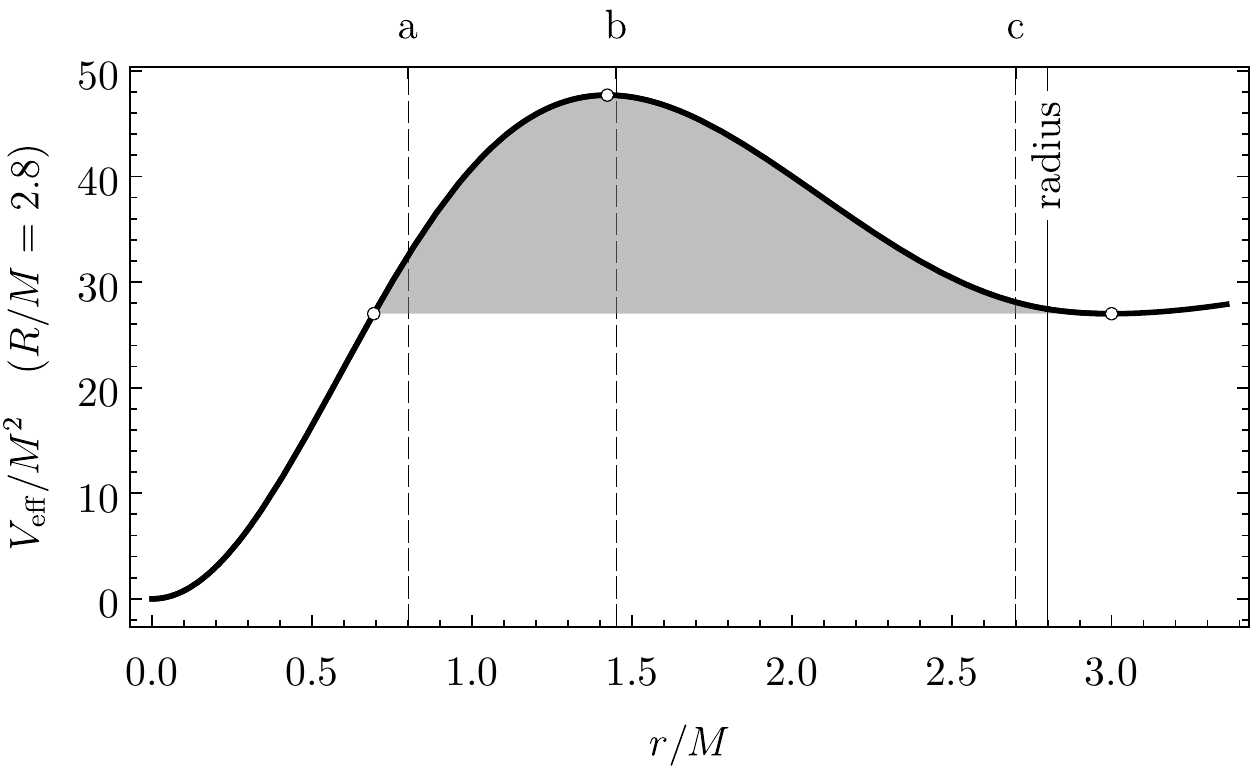}\\
     (a)\hspace*{-1em}\raisebox{-0.23\linewidth}{\includegraphics[width=0.25\linewidth]{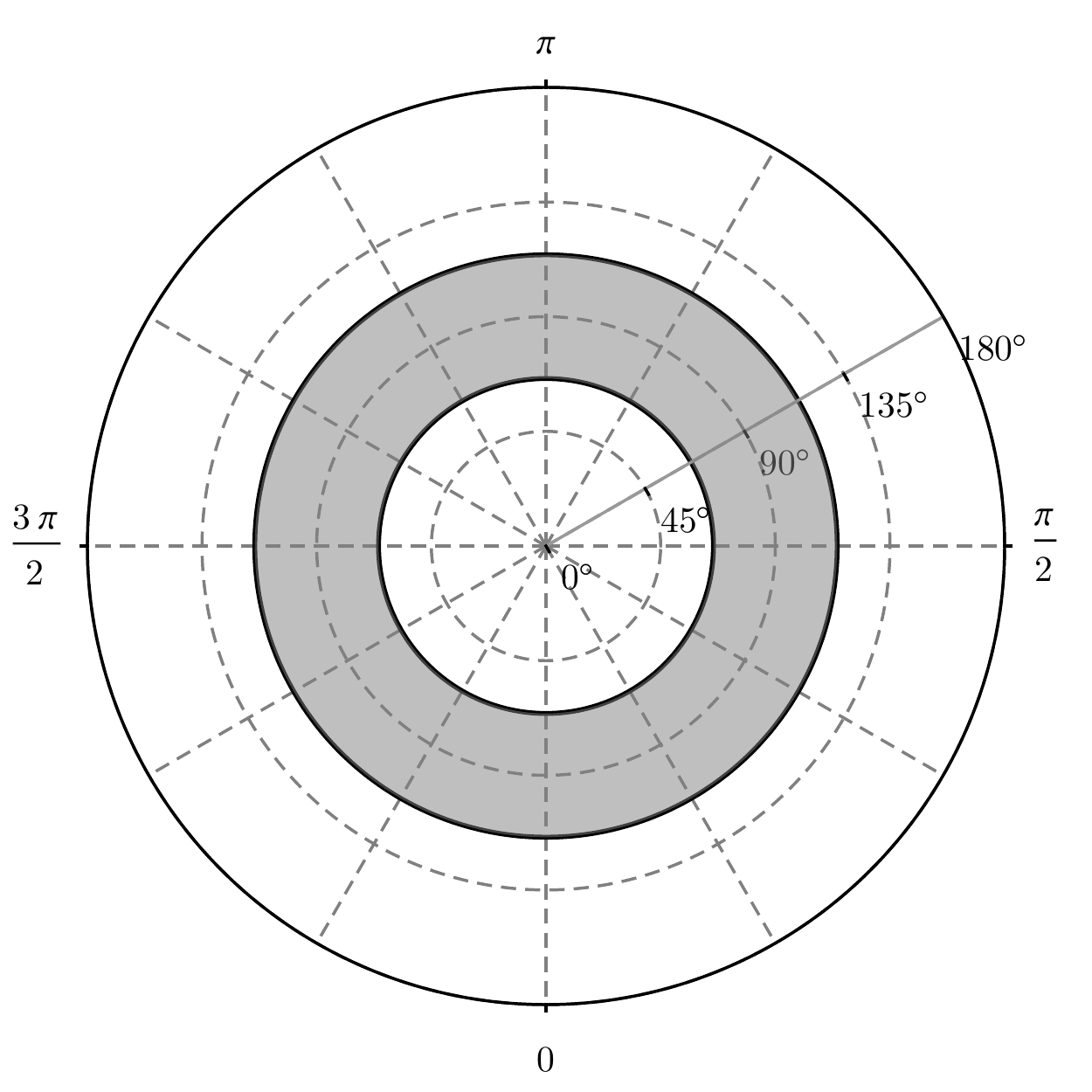}}\qquad%
     (b)\hspace*{-1em}\raisebox{-0.23\linewidth}{\includegraphics[width=0.25\linewidth]{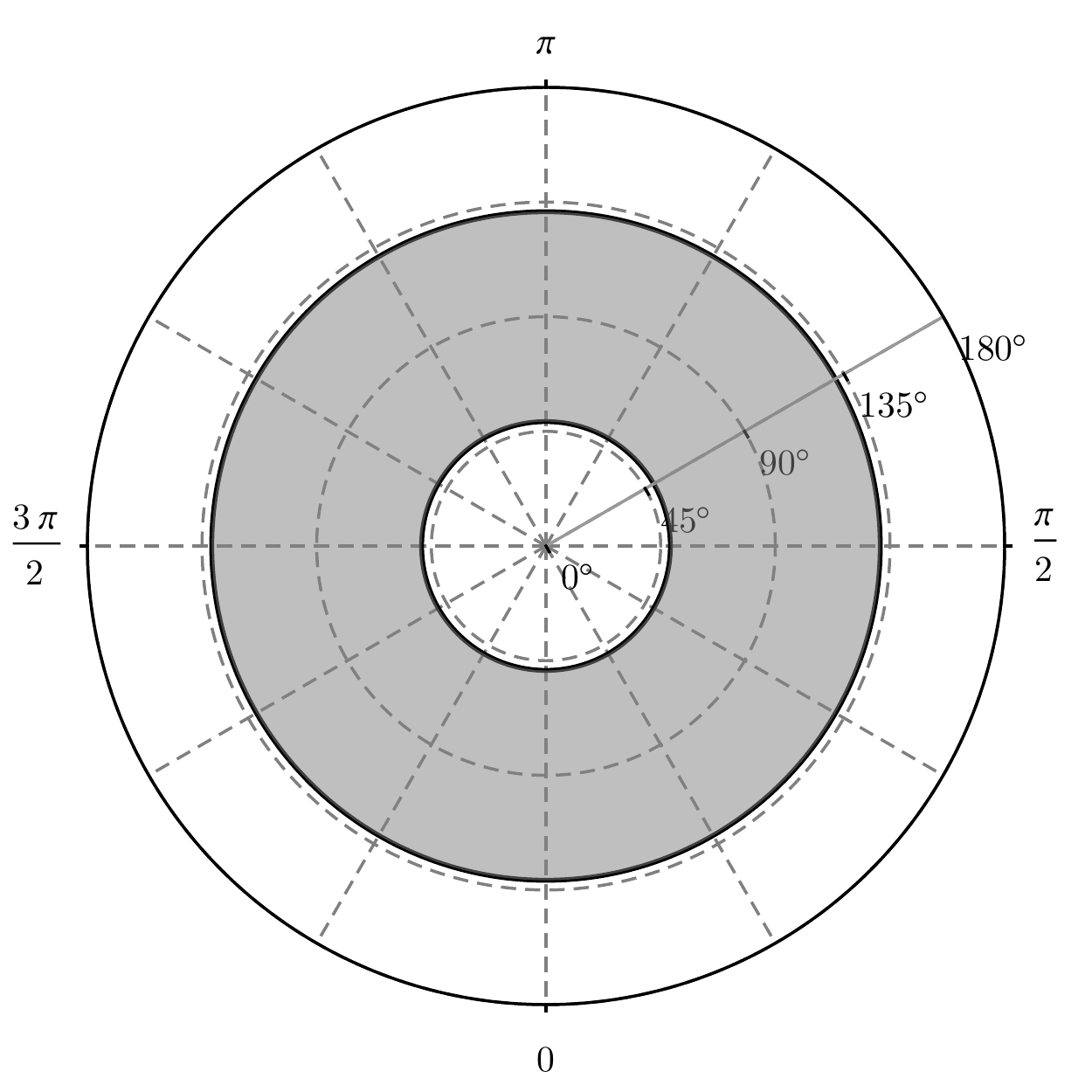}}\qquad%
     (c)\hspace*{-1em}\raisebox{-0.23\linewidth}{\includegraphics[width=0.25\linewidth]{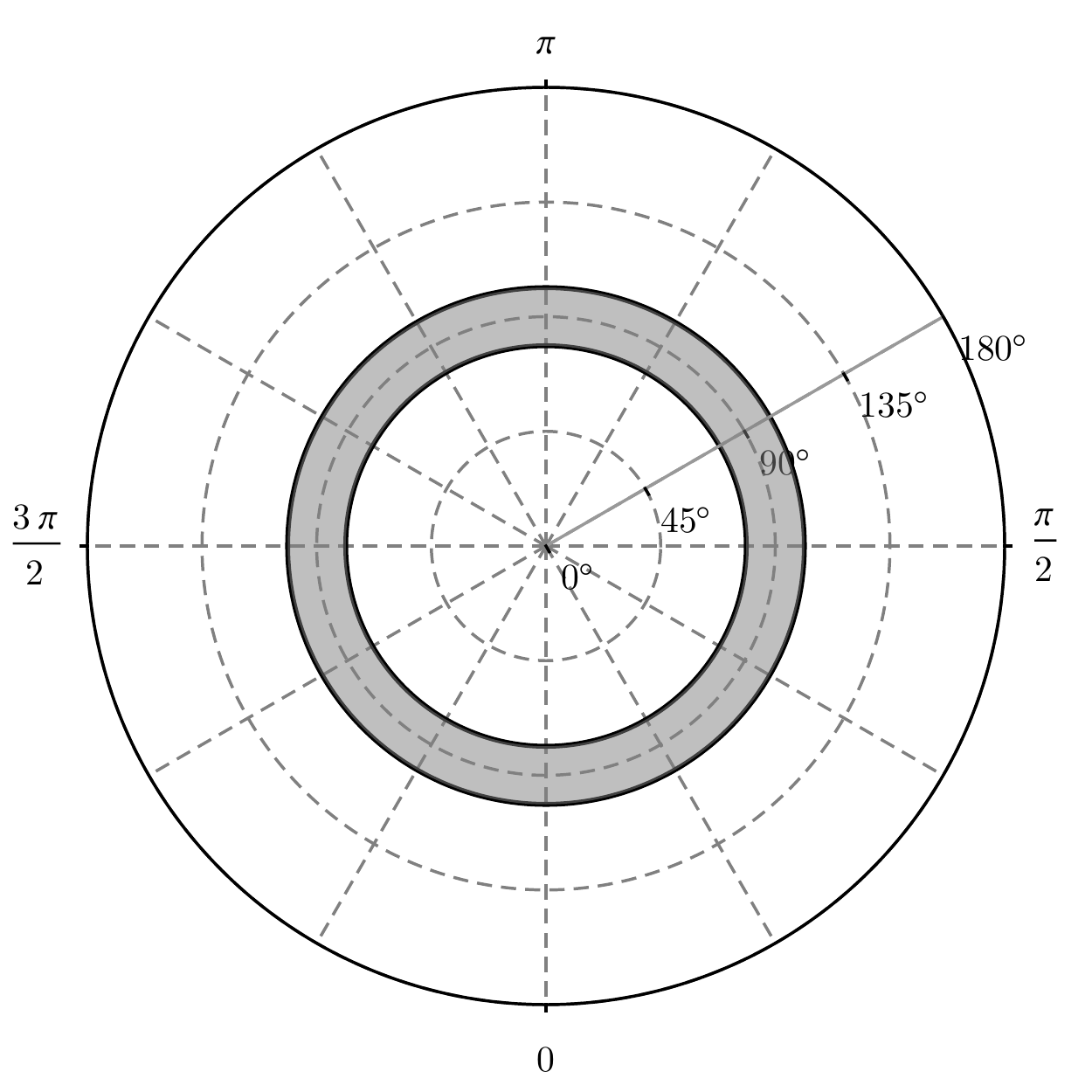}}
	\end{center}
	\caption{\label{figure6}The escaping/trapping cones corresponding to $r/M = \{0.8, 1.45, 2.7\}$ for the Tolman~VII configuration with $R/M = 2.8$. Above is shown the effective potential and marked positions indicate the locations where the cones are determined. The trapping zones are shaded.}
\end{figure*}
%%%%%%%%%%%%%%%%%%%%%%%%%%%%%%%%%%%%%%%%%%%%%%%%%%%%%%%%%%%%%%%%%%%%%%%%%%%%%%%%%%%%%%%%%%%%%%%%%%%%%%%%%%%%%%%%%%%%%%%%%%%%%%%%%%%%%%%%%%%%%%%%%%%%%%%%%%%%%%%%%%%%%%%%%%%%%%%%%%%%%%%%

In order to obtain the trapping (escape) cones in the observer (source) sky it is crucial that they are fully governed by the angles corresponding to the photon parameters defining the stable and unstable circular null geodesics. Therefore, it is sufficient to find the angles $\alpha_\mathrm{c(i)}$ corresponding to $\lambda_\mathrm{c(i)}$ of the stable circular null geodesic, and $\lambda_\mathrm{b(2))}$ of the unstable circular null geodesic. We thus relate the directional angles to the motion constant (impact parameters). Because of the spherical symmetry, we can consider for simplicity the equatorial null geodesics (when $\beta = 0$, or $\beta = \pi$, and $p^{(\theta)} = 0$). The directional angle $\alpha$ is then given by the relations
\begin{equation}
    \sin\alpha = \frac{p^{(\varphi)}}{p^{(t)}}\, ,\qquad \cos\alpha = \frac{p^{(r)}}{p^{(t)}}\, .
\end{equation}
The radial component of the null geodesic 4-momentum re\-ads
\begin{equation}
    p_r = \pm E\, \mathrm \ee^{\left(\Psi - \Phi\right)/2}\left(1 - \mathrm \ee^{\Phi}\,\frac{\lambda^2}{r^2}\right)^{1/2}\, ,
\end{equation}
and finally we can express the directional angle in the Tolman~VII spacetimes in the form (for simplicity $M = 1$)
\begin{align}
    \sin\alpha & = \ee^{\Phi/2}\,\frac{\lambda}{r} = \sqrt{1-5\mathcal{C}/3}\, \cos\left[C_\mathrm{a} + Y(r)\right] \,\frac{\lambda}{r}\, ,\\
    \cos\alpha & = \pm\sqrt{1 - \sin^2\alpha}\, .
\end{align}
To find the trapping (escape) cone in the region where the trapping is possible, which is defined by the extension of the effective potential barrier governed by its local extrema, we have to calculate the angles $\alpha_\mathrm{b(2)}$ corresponding to the impact parameter $\lambda=\lambda_\mathrm{b(2)}$ given by the relation
\begin{align}
    \cos\alpha_\mathrm{b(2)}(r,\mathcal{C}) &= \pm \sqrt{1 - \left( \ee^{\Phi(r)/2}\, \frac{\lambda_\mathrm{b(2)}}{r}\right)^2} \nonumber\\
                                            &= \pm \sqrt{1- \frac{V{}_\mathrm{eff}^\mathrm{int}(r_\mathrm{b(2)})}{V{}_\mathrm{eff}^\mathrm{int}(r)}}\, .
\end{align}
We have to separate the case $R/M \leq 3$ when $\lambda_\mathrm{b(2)}=3\sqrt{3}\, M$ in the external vacuum Schwarzschild spacetime, and the case $R/M > 3$ when $\lambda^{2}_\mathrm{b(2)} = V{}^\mathrm{int}_\mathrm{eff(min)}(\mathcal{C})$ in the internal Tolman~VII spacetime.

The trapping zone lies between the angles $\alpha_\mathrm{b(2)}$ related to the outward and inward radial direction, as shown in Fig.~\ref{figure5} where the trapping zone is light shaded, while the escape cone (zone) is dark shaded and null geodesics of this kind can escape to infinity even if originally radiated inwards.

%%%%%%%%%%%%%%%%%%%%%%%%%%%%%%%%%%%%%%%%%%%%%%%%%%%%%%%%%%%%%%%%%%%%%%%%%%%%%%%%%%%%%%%%%%%%%%%%%%%%%%%%%%%%%%%%%%%%%%%%%%%%%%%%%%%%%%%%%%%%%%%%%%%%%%%%%%%%%%%%%%%%%%%%%%%%%%%%%%%%%
\begin{figure*}[t]
    \includegraphics[width=0.48\linewidth]{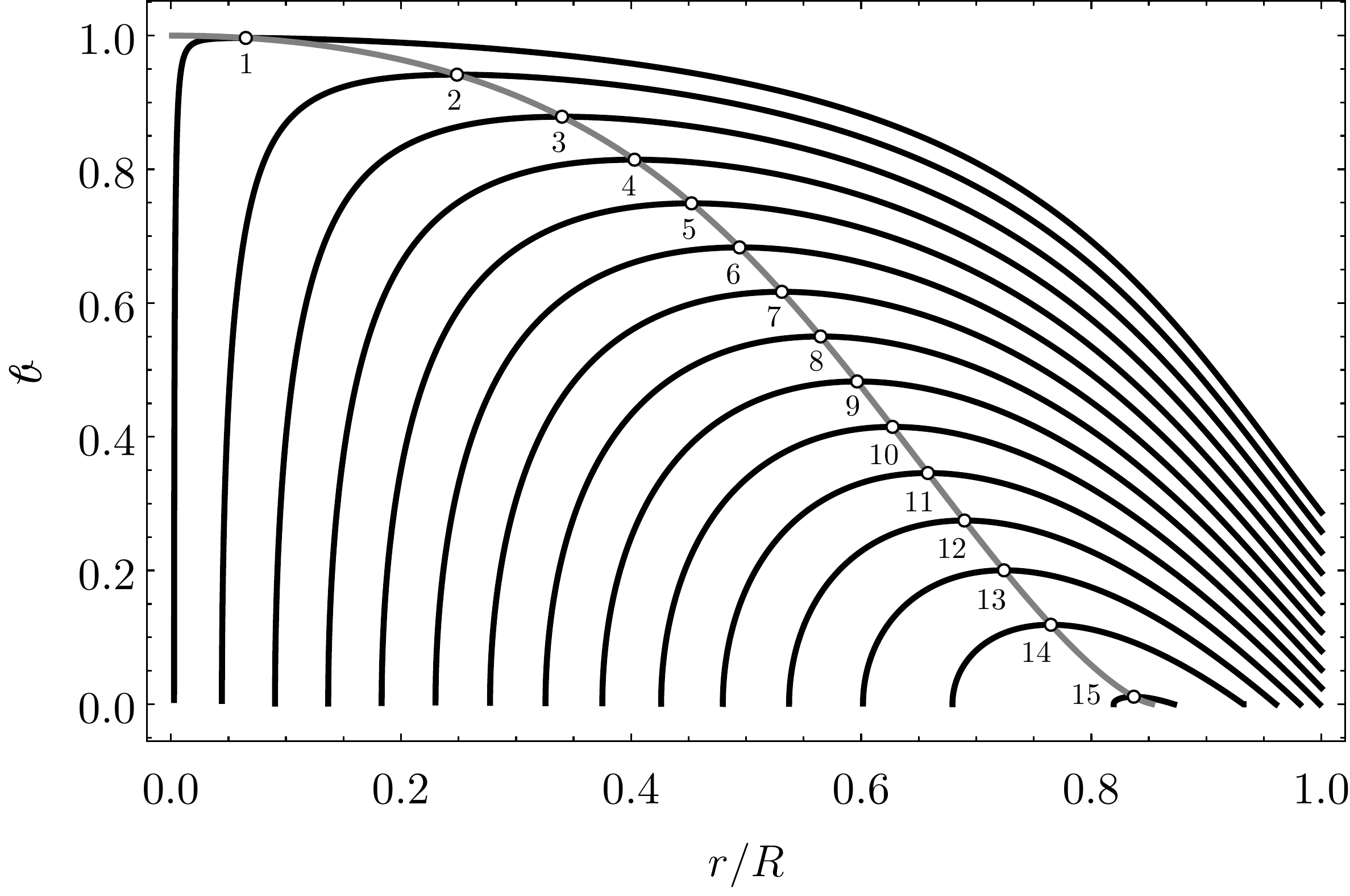}\hfill\includegraphics[width=0.48\linewidth]{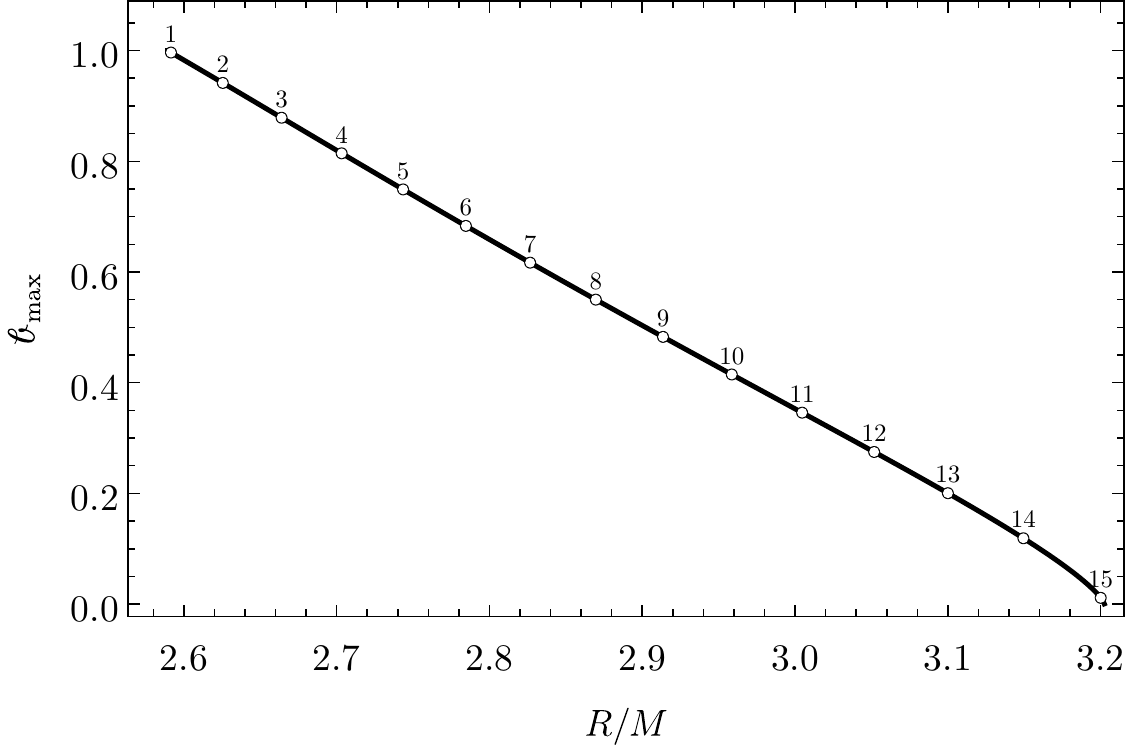}
    \caption{\label{figure7} The local trapping efficiency radial profile $\mathscr{b}(r,\mathcal{C})$ illustrated for several values of inverse compactness $R/M$ of the extremely compact Tolman~VII spacetimes (left plot). The parameter $R/M$ for each curve is given by the corresponding position of $\mathscr{b}_\mathrm{max}$ on the right plot which shows the local maxima of local trapping efficiency $\mathscr{b}_\mathrm{max}$ for the same values of the parameter $R/M$.}
\end{figure*}
%%%%%%%%%%%%%%%%%%%%%%%%%%%%%%%%%%%%%%%%%%%%%%%%%%%%%%%%%%%%%%%%%%%%%%%%%%%%%%%%%%%%%%%%%%%%%%%%%%%%%%%%%%%%%%%%%%%%%%%%%%%%%%%%%%%%%%%%%%%%%%%%%%%%%%%%%%%%%%%%%%%%%%%%%%%%%%%%%%%%%%%%

Recall that the trapping effects are relevant only in the extremely compact Tolman~VII spacetimes existing in the range $2.590 < R/M < 3.202$, and can occur only in the range of radii of these objects limited by $r_\mathrm{b(1)} < r < r_\mathrm{b(2)}$ for $R/M > 3$, and $r_\mathrm{b(1)} < r < R$, for $R/M < 3$. At any allowed radius $r$, the trapping occurs for the values of the impact parameter in the region $\lambda_\mathrm{b(2)} < \lambda < \lambda_\mathrm{t} \equiv \sqrt{V{}^\mathrm{int}_\mathrm{eff}(r)}$, while $\lambda_\mathrm{b(2)}$ corresponds to $\cos\alpha_\mathrm{b(2)}(r)$ and $\lambda_\mathrm{t}$ corresponds to $\cos\alpha_\mathrm{t} = 0$ ($\alpha = \pi/2$ corresponds to the turning point of the radial motion). At $r = r_\mathrm{c(i)}$, there is $\lambda_\mathrm{t} = \lambda_\mathrm{c(i)}$.

The extension of the trapping zone in the plane of angles $(\alpha, \beta)$ in dependence on the position of the source is presented for some representative values of the compactness in Fig.~\ref{figure6}. Because of the spacetime symmetry the trapping cones are symmetric relative to the center. \footnote{Note that in rotating spacetimes the symmetry of the trapping zone (cone) is lost as the motion depends on the sign of the impact parameter, as shown for the case of trapped null geodesics in Kerr spacetimes~\cite{Stu-Sch:2010:CQG,Stu-Cha-Sce:2018:EPJC}.}

%%%%%%%%%%%%%%%%%%%%%%%%%%%%%%%%%%%%%%%%%%%%%%%%%%%%%%%%%%%%%%%%%%%%%%%%%%%%%%%%%%%%%%%%%%%%%%%%%%%%%%%%%%%%%%%%%%%%%%%%%%%%%%%%%%%%%%%%%%%%%%%%%%%%%%%%%%%%%%%%%%%%%%%%%%%%%%%%%%%%%%%%
\section{Efficiency of neutrino trapping in extremely compact Tolman~VII spacetime}\label{SECTION4}
The trapping effect can be characterized by its efficiency which can be defined in both, the local sense taken at a given radius of the object, and the global sense considering the whole object~\cite{Stu-Tor-Hle:2009:CQG}. Denoting as $N_\mathrm{p}$, $N_\mathrm{t}$, and $N_\mathrm{e}$, respectively, the number of produced, trapped and escaped neutrinos, per unit time of distant static observers, the global trapping efficiency $\mathcal{B}_\mathrm{t}$ and global escaping efficiency $\mathcal{B}_\mathrm{e}$ are determined by the relations
\begin{equation}
\mathcal{B}_\mathrm{t} = \frac{N_\mathrm{t}(\mathcal{C})}{N_\mathrm{p}(\mathcal{C})}\, , \qquad \mathcal{B}_\mathrm{e} = \frac{N_\mathrm{e}(\mathcal{C})}{N_\mathrm{p}(\mathcal{C})}\, ,
\end{equation}
which satisfy $\mathcal{B}_\mathrm{t} + \mathcal{B}_\mathrm{e} = 1$. In order to find the global trapping efficiency, we have to find the local efficiency related to a fixed radius in the region where the trapping occurs. In the following, we consider for simplicity that the neutrinos are produced by sources emitting isotropically --- then the trapping will be locally governed solely by the spacetime geometry (for the case of anisotropic emission see \textit{e.g.}~\cite{Stu-Tor-Hle:2009:CQG}).

\subsection{Local efficiency of trapping}
To determine the local properties of the trapping effect, we introduce a local trapping efficiency coefficient $\mathscr{b}(r,\mathcal{C})$, defined at a given radius $r$ of the compact object as the ratio of the number of neutrinos emitted, from this radius, and trapped by the object $\mathrm{d}N_\mathrm{t}(r,\mathcal{C})$, to the number of neutrinos totally produced at this radius $\mathrm{d}N_\mathrm{p}(r,\mathcal{C})$.

Due to the isotropy of the radiation emitted by the local sources at the given radius, the number of escaping neutrinos is determined by the solid angle $\Omega_\mathrm{e}(\alpha_\mathrm{b(2)})$ given by
\begin{align}
    \Omega_\mathrm{e}(\alpha_\mathrm{b(2)}) &= \int_{0}^{\alpha_\mathrm{b(2)}}\int_{0}^{2\pi} \sin\alpha \, \mathrm{d}\alpha \, \mathrm{d}\varphi \nonumber\\
                                            &= 2\pi(1-\cos\alpha_\mathrm{b(2)})\, ,
\end{align}
while the number of produced neutrinos is the total solid angle $\Omega_\mathrm{p} = 4\pi$. Then the local escaping efficiency is given by the relation
\begin{align}
    \mathscr{e}(r,\mathcal{C}) \equiv \frac{\mathrm{d}N_\mathrm{e}(r)}{\mathrm{d}N_\mathrm{p}(r)} &= \frac{2\Omega(\alpha_\mathrm{b(2)}(r,\mathcal{C}))}{4\pi} \nonumber\\
                                                                                                  &= 1 - \cos\alpha_\mathrm{b(2)}(r,\mathcal{C})\, ,
\end{align}
and the local trapping coefficient, $\mathscr{b}(r,\mathcal{C}) + \mathscr{e}(r,\mathcal{C}) = 1$ reads
\begin{equation}
    \mathscr{b}(r,\mathcal{C}) \equiv \frac{\mathrm{d}N_\mathrm{b}(r)}{\mathrm{d}N_\mathrm{p}(r)} = \cos\alpha_\mathrm{b(2)}(r,\mathcal{C})\, .
\end{equation}
The resultant local trapping efficiency coefficient is presented in Fig.~\ref{figure7}.

The position of the local extrema of the local trapping efficiency function $\mathscr{b}(r,\mathcal{C})$, given by the condition $\partial\mathscr{b} / \partial r = 0$, coincides with the position of the stable circular null geodesic. The local extrema $\mathscr{b}_\mathrm{max}$ determines the position of the largest local trapping in a fixed Tolman~VII configuration. The dependence of $\mathscr{b}_\mathrm{max}$ on $R/M$ is plotted in Fig.~\ref{figure7}.

\begin{figure}[t]
    \centering
    \includegraphics[width=\linewidth]{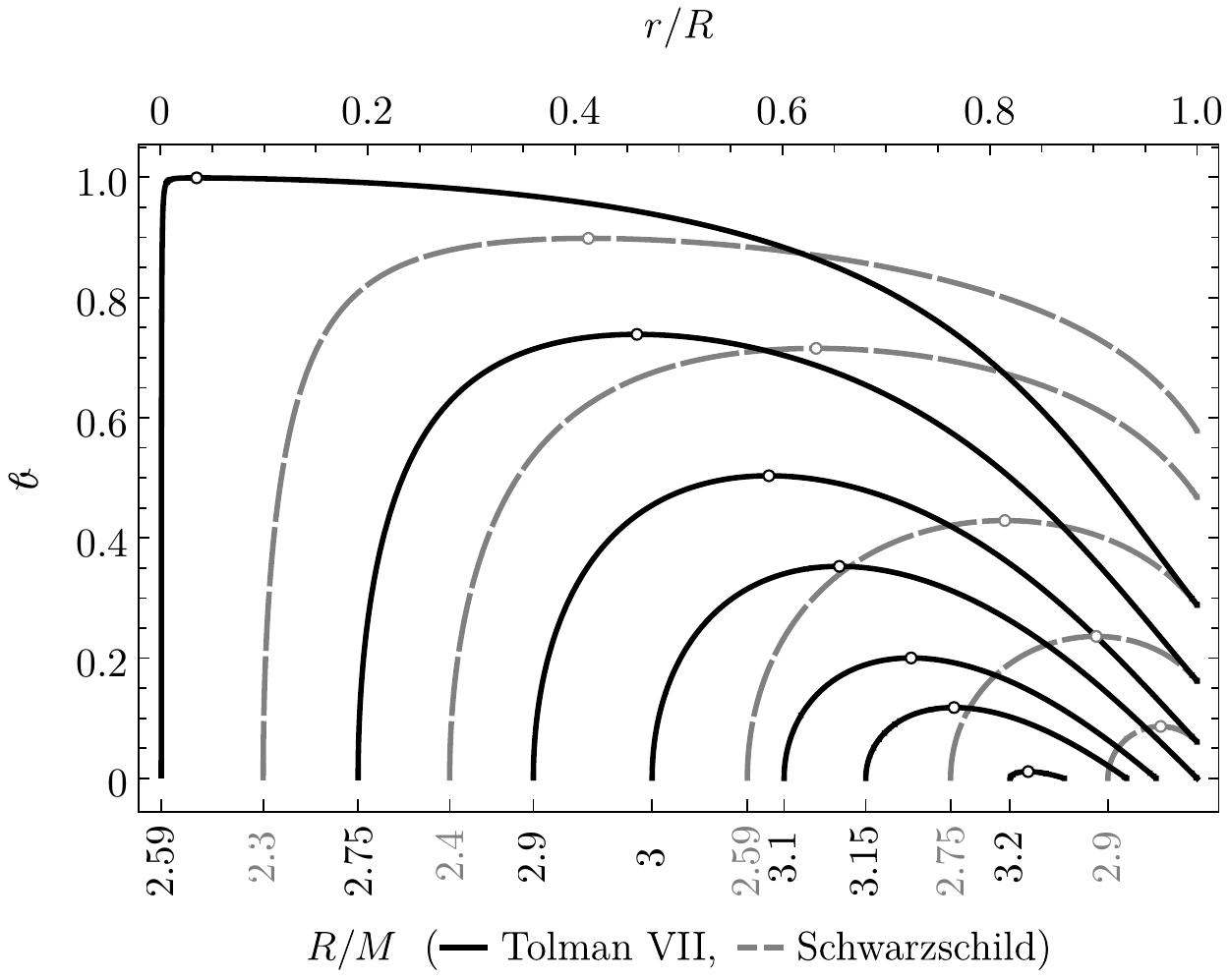}
    \caption{\label{figure8}Local trapping efficiency in extremely compact Tolman~VII spacetime (black) compared to local trapping efficiency in extremely compact Schwarzschild spacetime (dashed--gray).}
\end{figure}

For better insight into the nature of the trapping phenomena, in Fig.~\ref{figure8} we compare the results obtained for the Tolman~VII configurations to those related to the internal Schwarzschild spacetimes with uniformly distributed energy density~\cite{Stu-Tor-Hle:2009:CQG}, demonstrating some important differences. The most relevant one is that the radius of the extremely compact Tolman~VII solution can overcome the photosphere of the external vacuum Schwarzschild --- in this case the trapping region is not extending to the surface of the object, but its external boundary approaches the surface when $R \to 3 M$. We can also see that the difference in the local trapping efficiency profile for the maximally compact Tolman~VII spacetime is not large, but it shows a significant decrease near the central region of the object. On the other hand, for the Tolman~VII and interior Schwarzschild, with the same parameter $R/M$, the local trapping efficiency radial profile is higher for the Tolman~VII spacetimes, vanishing for $R/M = 3.202$, while for the internal Schwarzschild spacetimes the vanishing occurs for $R/M = 3$.

%%%%%%%%%%%%%%%%%%%%%%%%%%%%%%%%%%%%%%%%%%%%%%%%%%%%%%%%%%%%%%%%%%%%%%%%%%%%%%%%%%%%%%%%%%%%%%%%%%%%%%%%%%%%%%%%%%%%%%%%%%%%%%%%%%%%%%%%%%%%%%%%%%%%%%%%%%%%%%%%%%%%%%%%%%%%%%%%%%%%%%%%
%%%%%%%%%%%%%%%%%%%%%%%%%%%%%%%%%%%%%%%%%%%%%%%%%%%%%%%%%%%%%%%%%%%%%%%%%%%%%%%%%%%%%%%%%%%%%%%%%%%%%%%%%%%%%%%%%%%%%%%%%%%%%%%%%%%%%%%%%%%%%%%%%%%%%%%%%%%%%%%%%%%%%%%%%%%%%%%%%%%%%%%%

\subsection{Neutrino production}
In order to study the efficiency of the neutrino trapping in the extremely compact Tolman~VII spacetimes, in the global sense, reflecting both the cooling process and the total neutrino luminosity of the object, we have to discuss first the production rate of neutrinos and its properties that have to be well representative for our study of the Tolman~VII spacetimes.

The neutrino production is generally a complex process governed fully by the detailed structure of the configuration (\textit{e.g.}, a neutron star). The local neutrino production rate $I(r,\mathcal{A})$, considered at a given radius $r$, is determined by the relation
\begin{equation}
    I(r, \mathcal{A}) = \frac{\mathrm{d}N(r, \mathcal{A})}{\mathrm{d}\tau}\, ,
\end{equation}
where $\mathrm{d}N$ is the number of neutrino producing interactions at radius $r$, per element of proper time $\mathrm{d}\tau$ of the static observer located at the radius $r$; $\mathcal{A}$ denotes the full set of quantities governing the neutrino production rate, taken at the given radius. The number of interactions can be ex\-press\-ed in the form
\begin{equation}
    \mathrm{d}N(r, \mathcal{A}) = \mathrm{d}n(r) \, \Gamma(r, \mathcal{A}) \, \mathrm{d}V(r)\, ,
\end{equation}
where $\mathrm{d}n$ represents the number density of particles determining the neutrino production, $\Gamma$ denotes the neutrino production rate (governed by the temperature at the given radius) and $\mathrm{d}V$ denotes the proper volume element at the given radius; $\mathrm{d}n$ and $\Gamma$ are determined by the details of the matter in the extremely compact object, while $\mathrm{d}V$ is governed by the spacetime geometry. The production rate of neutrinos can be a very complex function of radius, being determined by all the complexities of the internal structure of the extremely compact object. However, it is rather meaningless to try to study the physical details of the compact object structure and related consequences on the neutrino production rate, in the case of the Tolman~VII solution, as it represents only a rough approximation of neutron stars modeled by realistic equations of state. In order to study the influence of the modification of the energy density profile in the Tolman~VII solution, it is quite reasonable and sufficient to assume that the neutrino production rate is determined by the energy density profile, as in the previous studies of the trapping effect in the internal Schwarzschild spacetimes~\cite{Stu-Tor-Hle:2009:CQG,Stu-Hla-Urb:2012:GRaG}; moreover, the energy density radial profile includes naturally, in an implicit way, the effect of temperature of the matter~\cite{Web:1999:BOOK}. The neutrino production rate is thus assumed in the form
\begin{equation}
    I(r) = \frac{\mathrm{d}N(r)}{\mathrm{d}\tau} \sim \rho(r)\, .
\end{equation}
We further assume as in~\cite{Stu-Tor-Hle:2009:CQG} that the neutrino radiation is locally isotropic so the efficiency of the trapping effect is given by the spacetime geometry (relative extension of the trapping cone) only.

Including the time-delay factor, we arrive to the neutrino production rate related to the distant static observers
\begin{equation}
    \mathcal{I}(r) = \frac{\mathrm{d}N(r)}{\mathrm{d}t} = I\, \ee^{\Phi(r)/2} .
\end{equation}
Finally, we find that the number of neutrinos generated at a given radius at the proper volume element $\mathrm{d}V$, per unit time of distant static observers, \textit{i.e.}, the local neutrino production rate, is determined by the relation
\begin{align}
    \mathrm{d}N_\mathrm{p}(r) &=  \mathcal{I}(r)\, \mathrm{dV}(r) = I(r)\, \ee^{\Phi(r)/2}\, 4\pi\, \ee^{\Psi(r)/2}\, r^2 \,\mathrm{d}r\, \nonumber\\
                              &\sim  \rho(r)\, 4\pi\, \ee^{\Phi(r)/2}\, \ee^{\Psi(r)/2}\, r^2 \, \mathrm{d}r \nonumber \\
                              & =  \rho_\mathrm{c}\left(1 - r^2/R^2\right) 4\pi\, \ee^{\Phi(r)/2}\, \ee^{\Psi(r)/2}\, r^2 \, \mathrm{d}r\, .
\end{align}
The global neutrino production rate is then determined by the integration across the whole compact object
\begin{align}
    N_\mathrm{p} &=  4\pi \int_{0}^{R} I(r)\, \ee^{\Phi(r)/2}\, \ee^{\Psi(r)/2}\, r^2 \, \mathrm{d}r \nonumber\\
                 &=  4\pi \int_{0}^{R} \rho_\mathrm{c}\left(1 - r^2/R^2\right) \ee^{\Phi(r)/2}\, \ee^{\Psi(r)/2}\, r^2 \, \mathrm{d}r\, .
\end{align}
The global rate of the neutrino trapping is determined under the assumption of the isotropy of the emitted neutrino flow given by the relation

\begin{multline}
    N_\mathrm{t} =  4\pi \int_{r_\mathrm{b(1)}}^{\min\{R,\,r_\mathrm{b(2)}\}} \rho_\mathrm{c}\left(1 - r^2/R^2\right) \mathscr{b}(r,R)\\ \times \ee^{\Phi(r)/2} \ee^{\Psi(r)/2} r^2\, \mathrm{d}r\, .
\end{multline}

Now, we are able to study the trapping of neutrinos and the global trapping efficiency in dependence on the spacetime parameters. We assume sources emitting isotropically the neutrinos following the null geodesics~\cite{Stu-Tor-Hle:2009:CQG}.

\subsection{Global trapping efficiency for total neutrino luminosity}
The coefficient of the global trapping related to total neutrino luminosity of the object characterizes the trapping phenomenon integrated across the whole trapping region, related to the radiating object. Thus, we consider the amount of neutrinos radiated along null geodesics by the object, per unit of distant observer time, and determine the part of these radiated neutrinos that remain trapped by the radiating object. Details of the derivation of the global trapping coefficient are presented in~\cite{Stu-Tor-Hle:2009:CQG}, and we apply them here using again the basic assumption that the locally defined radiation intensity is proportional to the energy density of the object matter, being thus distributed due to the quadratic radial profile of the Tolman~VII energy density. We thus again assume the emissivity directly related to the energy density of the object, enabling also easy comparison to the results of the study of the internal Schwarzschild spacetime. Of course, we could make a detailed calculation of the emissivity, using relevant physical conditions in the interior of neutron stars, and taking into account both the local particle density (proportional to the rest energy density), the temperature of matter, and all the details of the physics of neutrino emission~\cite{Gle:2000:BOOK,Web:1999:BOOK}. However, we decide to use the simple assumption of emissivity related to the energy density, as this assumption, in a reasonable measure, incorporates both the influence of the particle density of radiating matter and its temperature.

The global luminosity trapping effects are thus manifested by the global luminosity trapping efficiency coefficient $\mathcal{B}_\mathrm{L}$ defined by a suitable modification of the relation presented in~\cite{Stu-Tor-Hle:2009:CQG}
\begin{equation}
    \mathcal{B}_\mathrm{L} = \frac{\displaystyle\int_{r_\mathrm{b(1)}}^{\min{\{R,\,r_\mathrm{b(2)}\}}}\rho(r)\,\mathscr{g}(r,\mathcal{C})\,\mathscr{b}(r,\mathcal{C}) \, r^2 \, \mathrm{d}r}{\displaystyle\int_{0}^{R}\rho(r)\,\mathscr{g}(r,\mathcal{C}) \, r^2 \, \mathrm{d}r}\, ,
\end{equation}
where $\rho(r)$ is given by Eq.~\ref{eq13}, and
\begin{equation}
    \mathscr{g}(r,\mathcal{C}) = \ee^{(\Phi + \Psi)/2} = \sqrt{\frac{C_1 \cos^2\left(\phi_\mathrm{T}(r)\right)}{1 - \mathcal{C}\left(r/R\right)^2(5 - 3\left(r/R\right)^2)}}\, .
\end{equation}
The upper limit of the integral is $R$, if $R/M \leq 3$, and $r_\mathrm{b(2)}$, if $R/M > 3$. Contrary to the case of the internal Schwarzschild spacetimes, the integration must be carried out numerically. The results obtained for the extremely compact Tolman~VII spacetimes are presented in Fig.~\ref{figure9}, where for comparison we present also the results obtained for the extremely compact internal Schwarzschild spacetimes. Recall that the modeling of the cooling process has to be done using a Monte Carlo method taking into account the finiteness of the mean free path of neutrinos and possible scattering of trapped neutrinos that could cause change of their impact parameter and eventual escape. Of course, in such a case one have to distinguish the case of the interior and exterior trapped neutrinos~\cite{Stu-Hla-Urb:2012:GRaG}.

\subsection{Global trapping efficiency for neutrino cooling}
The influence of the neutrino trapping effect on the cooling process of the compact object can be effectively shown by the local trapping coefficient $\mathscr{b}(r,\mathcal{C})$ presented above --- clearly, this coefficient indicates that the cooling is most efficiently influenced near the location of the stable circular null geodesic where the efficiency is highest.

\begin{figure*}[t]
    \centering
    \includegraphics[width=.6\linewidth]{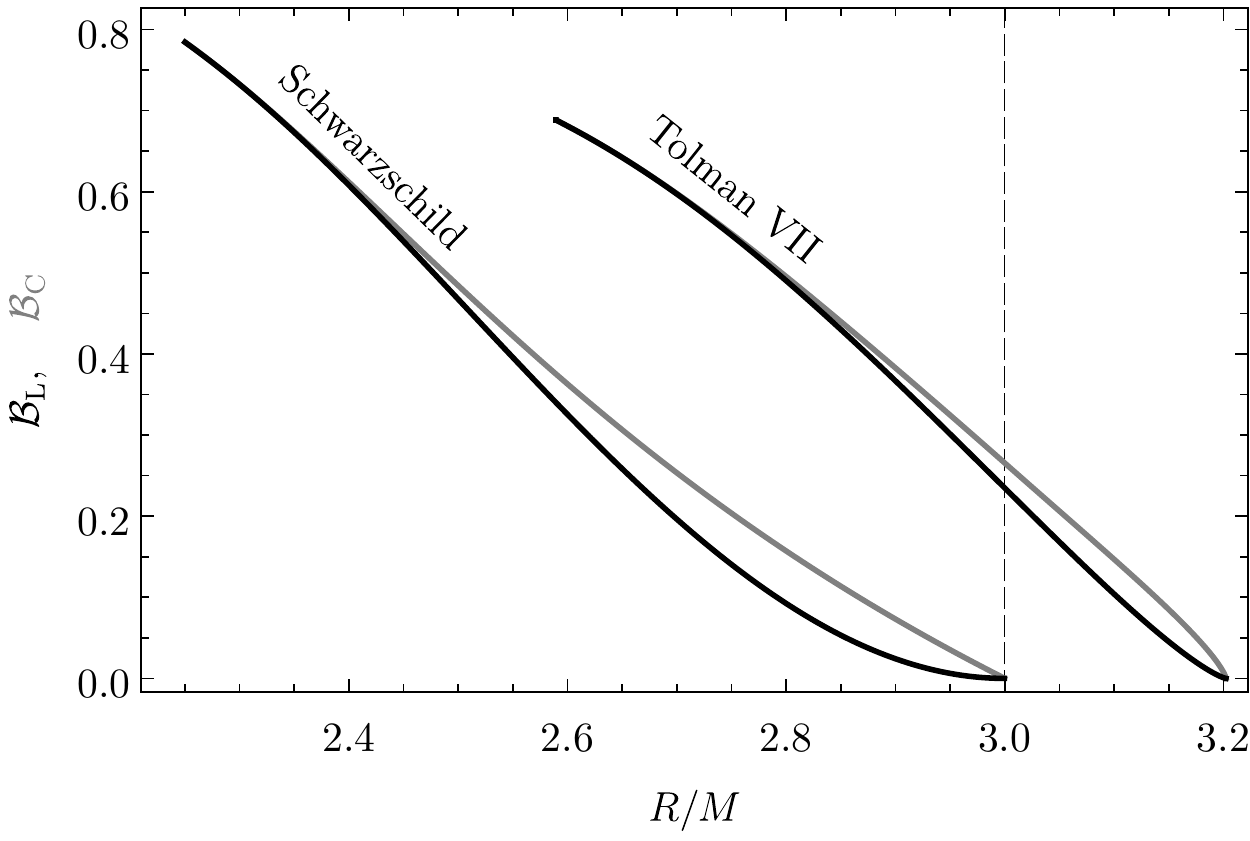}
    \caption{\label{figure9} The dependence of luminosity global trapping efficiency coefficient $\mathcal{B}_\mathrm{L}$ (black) and cooling global trapping coefficient $\mathcal{B}_\mathrm{C}$ (grey) for Schwarzschild and Tolman~VII star.}
\end{figure*}

However, the cooling efficiency can be represented also by a global trapping coefficient restricted exclusively to the active zone of the trapping that can be defined in a way closely related to the definition of the global trapping coefficient for total luminosity. Let us define the cooling global trapping coefficient by the relation
\begingroup
\allowdisplaybreaks
\begin{align}
    \mathcal{B}_\mathrm{C} & = \frac{\displaystyle\int_{r_\mathrm{b(1)}}^{\min{\{R,\,r_\mathrm{b(2)}\}}}\rho(r)\mathscr{g}(r,\mathcal{C})\,\mathscr{b}(r,\mathcal{C})\, r^2 \, \mathrm{d}r}{\displaystyle\int_{r_\mathrm{b(1)}}^{\min{\{R,\,r_\mathrm{b(2)}\}}}\rho(r)\,\mathscr{g}(r,\mathcal{C}) \, r^2 \, \mathrm{d}r} \nonumber\\
                           & = \frac{\displaystyle\int_{r_\mathrm{b(1)}}^{\min{\{R,\,r_\mathrm{b(2)}\}}}\left(1 - r^2/R^2\right)\mathscr{g}(r,\mathcal{C})\,\mathscr{b}(r,\mathcal{C})\,  r^2 \, \mathrm{d}r}{\displaystyle\int_{r_\mathrm{b(1)}}^{\min{\{R,\,r_\mathrm{b(2)}\}}}\left(1 - r^2/R^2\right)\mathscr{g}(r,\mathcal{C})\, r^2 \, \mathrm{d}r}\, .
\end{align}
\endgroup
In contrast with the definition of the global coefficient of total luminosity, now the interval of integration is the same in the numerator and denominator, all the functions occurring in the global coefficient definitions are the same in both cases. The results of the integration are presented in Fig.~\ref{figure9}, including the case of the internal Schwarzschild spacetimes for comparison. We notice that the global trapping coefficient of cooling slightly exceeds those related to the total luminosity, but their difference is suppressed with decreasing parameter $R/M$ for both cases.

%%%%%%%%%%%%%%%%%%%%%%%%%%%%%%%%%%%%%%%%%%%%%%%%%%%%%%%%%%%%%%%%%%%%%%%%%%%%%%%%%%%%%%%%%%%%%%%%%%%%%%%%%%%%%%%%%%%%%%%%%%%%%%%%%%%%%%%%%%%%%%%%%%%%%%%%%%%%%%%%%%%%%%%%%%%%%%%%%%%%%%%%%
%%%%%%%%%%%%%%%%%%%%%%%%%%%%%%%%%%%%%%%%%%%%%%%%%%%%%%%%%%%%%%%%%%%%%%%%%%%%%%%%%%%%%%%%%%%%%%%%%%%%%%%%%%%%%%%%%%%%%%%%%%%%%%%%%%%%%%%%%%%%%%%%%%%%%%%%%%%%%%%%%%%%%%%%%%%%%%%%%%%%%%%%%

%%%%%%%%%%%%%%%%%%%%%%%%%%%%%%%%%%%%%%%%%%%%%%%%%%%%%%%%%%%%%%%%%%%%%%%%%%%%%%%%%%%%%%%%%%%%%%%%%%%%%%%%%%%%%%%%%%%%%%%%%%%%%%%%%%%%%%%%%%%%%%%%%%%%%%%%%%%%%%%%%%%%%%%%%%%%%%%%%%%%%%%%%
%%%%%%%%%%%%%%%%%%%%%%%%%%%%%%%%%%%%%%%%%%%%%%%%%%%%%%%%%%%%%%%%%%%%%%%%%%%%%%%%%%%%%%%%%%%%%%%%%%%%%%%%%%%%%%%%%%%%%%%%%%%%%%%%%%%%%%%%%%%%%%%%%%%%%%%%%%%%%%%%%%%%%%%%%%%%%%%%%%%%%%%%%

\section{Conclusions}\label{SECTION5}
In the present study we considered the trapping of null geo\-de\-sics in relation to trapping of neutrinos in the extremely compact Tolman~VII spacetimes that are considered by some authors as the best representation of neutron stars given by an exact solution of the Einstein gravity~\cite{Nea-Ish-Lak:2001:PRD,Jia-Yag:2019:PRD,Jia-Yag:2020:PRD}. It is important that these spacetimes can be extremely compact (contain trapped null geodesics) even for $R/M \sim 3.2$ close to the values of observed neutron stars.

In our study we applied simplification of the isotropic emission of neutrinos by all sources in the compact object (as in~\cite{Stu-Tor-Hle:2009:CQG}) and assume also its linear dependence on the energy density of the object that is chosen in the Tolman~VII solutions to be quadratic, giving thus a simple generalization of the physically unrealistic uniform distribution of the energy density in the internal Schwarzschild spacetime.
We have found that the local trapping coefficient demonstrates behavior similar to those of the internal Schwarzschild spa\-ce\-ti\-mes, having maximal points located at the position of the stable circular null geodesic of the spacetime. For the objects of the same parameter $R/M$, we have found radial profiles of the local trapping coefficients in Tolman~VII to be located above those of the interior Schwarzschild. We can see that for the Tolman~VII solution with $R/M = 3$ (when trapping vanishes in the internal Schwarzschild spacetimes) about $0.35$ part of the emitted neutrinos is trapped near the location of the stable circular null geodesics indicating thus a strong possible role of trapped neutrinos on the cooling process.

Similar behavior is observed also in the case of the global trapping coefficients, both for the total luminosity, and the cooling process. For the spacetimes with the same $R/M$, both the global coefficients are significantly higher in the case of the Tolman~VII spacetimes. For example, at $R/M = 3$ representing the limit on the existence of extremely compact internal Schwarzschild solutions, we found for the Tolman~VII solution $\mathcal{B}_\mathrm{L} \sim 0.24$ and $\mathcal{B}_\mathrm{C} \sim 0.27$. Generally, the global cooling efficiency is slightly higher than the global luminosity coefficient and their difference increases with increasing value of the parameter $R/M$.

Our results indicate the following important conclusion --- for the physically realistic Tolman~VII solutions that could well reflect some important properties of neutron stars~\cite{Jia-Yag:2020:PRD}, the trapping of neutrinos could be relevant in physically realistic situations, demonstrating significant influence especially in the cooling process of the neutron stars having a cumulative character with possible effect on their structure, and smaller effect on their total luminosity. We also expect on the base of our previous results with rotating internal Schwarzschild spacetimes~\cite{Vrb-Urb-Stu:2020:EPJC} that rotational effect could lead to further enhancement of the role of the trapping in both the luminosity and cooling process, enabling it for slow\-ly rotating Tolman~VII objects with $R/M > 3.2$.

%%%%%%%%%%%%%%%%%%%%%%%%%%%%%%%%%%%%%%%%%%%%%%%%%%%%%%%%%%%%%%%%%%%%%%%%%%%%%%%%%%%%%%%%%%%%%%%%%%%%%%%%%%%%%%%%%%%%%%%%%%%%%%%%%%%%%%%%%%%%%%%%%%%%%%%%%%%%%%%%%%%%%%%%%%%%%%%%%%%%%%%%%
%%%%%%%%%%%%%%%%%%%%%%%%%%%%%%%%%%%%%%%%%%%%%%%%%%%%%%%%%%%%%%%%%%%%%%%%%%%%%%%%%%%%%%%%%%%%%%%%%%%%%%%%%%%%%%%%%%%%%%%%%%%%%%%%%%%%%%%%%%%%%%%%%%%%%%%%%%%%%%%%%%%%%%%%%%%%%%%%%%%%%%%%%

\begin{acknowledgement}
The authors acknowledge the institutional support of the Institute of Physics at the Silesian University in Opava.
\end{acknowledgement}

\bibliographystyle{spphys}
%\bibliography{references}

%%%%%%%%%%%%%%%%%%%%%%%%%%%%%%%%%%%%%%%%%%%%%%%%%%%%%%%%%%%%%%%%%%%%%%%%%%%%%%%%%%%%%%%%%%%%%%%%%%%%%%%%%%%%%%%%%%%%%%%%%%%%%%%%%%%%%%%%%%%%%%%%%%%%%%%%%%%%%%%%%%%%%%%%%%%%%%%%%%%%%%%%%
%%%%%%%%%%%%%%%%%%%%%%%%%%%%%%%%%%%%%%%%%%%%%%%%%%%%%%%%%%%%%%%%%%%%%%%%%%%%%%%%%%%%%%%%%%%%%%%%%%%%%%%%%%%%%%%%%%%%%%%%%%%%%%%%%%%%%%%%%%%%%%%%%%%%%%%%%%%%%%%%%%%%%%%%%%%%%%%%%%%%%%%%%
\end{document}